\documentclass[12pt,number]{elsarticle}

\usepackage{amssymb}
\usepackage{amsmath}
\usepackage{hyperref}
\usepackage{booktabs}
\usepackage{nicefrac}
\usepackage{listings}
\usepackage{xcolor}
\usepackage{graphicx}
\usepackage{subcaption}
\usepackage{float}
\usepackage{tikz}
\usepackage{enumitem}
\usepackage{siunitx}
\usepackage{multirow}
\usepackage{microtype}
\usepackage{orcidlink}
\usepackage{algorithm2e}
\usepackage{makecell}
\usepackage{pgfplots}
\pgfplotsset{compat=1.18}
\usetikzlibrary{arrows.meta, positioning, fit, backgrounds, calc,
                shapes.geometric}
\usepgflibrary{arrows.meta}
\usepackage{fontawesome5}

\definecolor{extblue}{RGB}{0, 102, 204}
\definecolor{webgreen}{RGB}{0, 128, 0}
\definecolor{catpurple}{RGB}{128, 0, 128}
\definecolor{sharedgray}{RGB}{169, 169, 169}
\definecolor{mcporg}{RGB}{204,102,0}
\definecolor{apirust}{RGB}{178,34,34}
\definecolor{codebg}{HTML}{F8F8F8}
\definecolor{codeframe}{HTML}{DDDDDD}

\begin{document}

\setlength{\emergencystretch}{3em}
\sloppy
\renewcommand{\labelenumii}{\arabic{enumi}.\arabic{enumii}}

\begin{frontmatter}

\title{Tokalator: A Context Engineering Toolkit for {Artificial Intelligence} Coding
       Assistants}

\author[1]{Vahid Farajijobehdar~\orcidlink{0000-0002-9153-0765}}
\ead{vahid.farajijobehdar@kariyer.net}

\author[1]{\.{I}lknur K\"{o}seo\u{g}lu Sar\i~\orcidlink{0000-0001-6153-6245}}
\ead{ilknur.koseoglu@kariyer.net}

\author[2]{Naz\i m Kemal \"{U}re~\orcidlink{0000-0003-2660-2141}}
\ead{ure@stanford.edu}

\author[3]{Engin Zeydan\corref{cor1}~\orcidlink{0000-0003-3329-0588}}
\ead{engin.zeydan@cttc.cat}

\cortext[cor1]{Corresponding author}

\affiliation[1]{
  organization={Kariyer.net, R\&D Center},
  city={Istanbul},
  country={Turkey}
}

\affiliation[2]{
  organization={Stanford University and iLab},
  city={Stanford, CA},
  country={U.S.A.}
}

\affiliation[3]{
  organization={Centre Tecnol\`ogic de Telecomunicacions de
                Catalunya (CTTC/CERCA)},
  city={Castelldefels},
  country={Spain}
}

\begin{abstract}
Artificial Intelligence (AI)-assisted coding environments operate within finite context windows of 128,000-1,000,000 tokens (as of early 2026), yet existing tools offer limited support for monitoring and optimizing token consumption. As developers open multiple files, model attention becomes diluted and Application Programming Interface (API) costs increase in proportion to input and output as conversation length grows. Tokalator is an open-source context-engineering toolkit that includes a VS Code extension with real-time budget monitoring and 11 slash commands; nine web-based calculators for Cobb-Douglas quality modeling, caching break-even analysis, and $O(T^2)$ conversation cost proofs; a community catalog of agents, prompts, and instruction files; an MCP server and Command Line Interface (CLI); a Python econometrics API; and a PostgreSQL-backed usage tracker. The system supports 17 Large Language Models (LLMs) across three providers (Anthropic, OpenAI, Google) and is validated by 124 unit tests. An initial deployment on the Visual Studio Marketplace recorded 313 acquisitions with a 206.02\% conversion rate as of v3.1.3. A structured survey of 50 developers across three community sessions indicated that instruction-file injection and low-relevance open tabs are among the primary invisible budget consumers in typical AI-assisted development sessions.

\end{abstract}

\begin{highlights}
\item VS~Code extension tracks token budgets for 17 LLM models across five cost categories
\item Five-signal scorer identifies distractor tabs; evaluated at six F1 thresholds
\item Closed-form caching break-even ($n^*=2$), $O(T^2)$ cost growth, Cobb--Douglas optimization
\item MCP server + CLI provide real Claude Byte Pair Encoding (BPE) token counting for Claude Code agents
\item 124 unit tests verify mathematical models; +50 out of 220 developers gave qualitative feedback during interactive sessions
\end{highlights}


\end{frontmatter}

\section{Introduction and Motivation}
\label{sec:intro}

Modern AI coding assistants such as GitHub Copilot (VS Code), Claude Code, and Cursor, help programmers develop code efficiently by connecting their integrated development environment (IDE) to large language models (LLMs) with context windows of 128,000 to 1,000,000 tokens (as of early 2026). Aubakirova et al.~\cite{aubakirova2025state}, drawing on over 100 trillion tokens of real-world interactions on the OpenRouter platform, document a structural shift: average prompt length grew nearly fourfold between 2024 and 2025 (from $\approx$1,500 to $>$6,000 tokens), driven by agentic workflows and reasoning-intensive tasks. Robbes et al.~\cite{robbes2026agentic} confirmed widespread adoption, finding 15.85--22.60\% of 129,134 active GitHub projects now using AI coding agents, sessions that consume far more tokens per interaction than traditional completions.  Despite ever-larger context windows, developers lack visibility into \emph{how} their context budget is consumed. Every open tab, system prompt, instruction file, and conversation turn contributes silently to this budget.
For example, at Anthropic's Claude Opus~4.6 pricing of \$5.00/MTok input and \$25.00/MTok output~\cite{anthropic2026pricing}, a single 200,000-token prompt costs \$1.00 for input alone.
Existing tools address only fragments of this problem.
\texttt{tiktoken}~\cite{sennrich2016bpe} and Anthropic's tokenizer count tokens offline but have trivial IDE integration, no cross-provider support, and no cost modeling.
Claude Code exposes \texttt{/context} (current context size) and \texttt{/cost} (session spend) as CLI commands, but these are Claude-specific, terminal-only, and provide no per-file breakdown, no cross-provider comparison, and no caching or conversation-strategy
analysis. VS~Code~v1.110~\cite{vscode2026v110}, released concurrently, added a native context indicator and \texttt{/compact} slash command, but both
are scoped to Copilot sessions and do not extend to other providers or expose economic models.
Han et al.~\cite{han2024talebudget} showed token budgets can be
enforced at the reasoning level without quality loss, but no IDE tool previously exposed this control in a cross-provider, cost-modelling form.
Bergemann et al.~\cite{bergemann2025menu} formalised LLM output quality as a Cobb--Douglas production function but did not implement a developer-facing tool.
Without integrated tooling, three compounding problems arise:

\begin{enumerate}[leftmargin=*]
  \item \textit{Attention dilution}: irrelevant files compete with
    relevant ones for the context window, reducing model output
    quality~\cite{bergemann2025menu,hong2025contextrot,liu2024lost}.
\item \textit{Cost rise}: conversations sending full history at every
turn grow at $O(T^2)$ cumulative cost, where $T$ is the total number
of conversation turns. At turn~$t$ the model receives $S + t(u + a)$
input tokens (system prompt~$S$, average user tokens~$u$, average
assistant tokens~$a$), so total input cost is:
\begin{equation}
  \sum_{t=1}^{T}\bigl[S + t(u+a)\bigr]
  = ST + \frac{T(T+1)}{2}(u+a) \in O(T^2).
  \label{eq:cost-rise}
\end{equation}
  \item \textit{Context rot}: after 20+ turns, stale context degrades
    model accuracy; Hong et al.~\cite{hong2025contextrot} showed this
    degradation is uneven across LLM models, appears even on simple
    retrieval tasks, and worsens when distractor content is present.
\end{enumerate}

This paper addresses three research questions (RQ):
\label{sec:intro_rqs}

\begin{description}
  \item[\textbf{RQ1}] \textit{What are the primary sources of token
    budget consumption in a typical AI-assisted development session,
    and can they be made visible to developers in real time?}

  \item[\textbf{RQ2}] \textit{Can a lightweight syntactic relevance scorer reliably identify \emph{distractor tabs},open IDE files that contribute tokens to the context window but are not relevant to the current coding task, that developers agree should be removed?}

  \item[\textbf{RQ3}] \textit{Can formal economic models (Cobb--Douglas production functions, caching break-even analysis, conversation cost projections) be implemented as practical developer tools without requiring access to proprietary model internals?}
\end{description}

Tokalator (a portmanteau of ``token'' and ``calculator'')\footnote{%
  Source code: \url{https://github.com/vfaraji89/tokalator};
  VS~Code Marketplace:
  \url{https://marketplace.visualstudio.com/items?itemName=vfaraji89.tokalator};
  Web platform: \url{https://tokalator.wiki/}.}
is a VS~Code extension that monitors context budget consumption
in real time and helps developers reduce token waste and API costs.
Its relevance scorer applies weighted rules over five syntactic
signals to identify low-relevance open tabs;
its economic calculators translate formal LLM cost models into
concrete cost estimates that developers can act on directly.
The contributions are framed to address the specific gaps:

\begin{enumerate}[leftmargin=*]

  \item \textbf{In-IDE, five-category context budget monitor.}
  Existing tokenizers (\texttt{tiktoken}, Anthropic's offline API) count aggregate tokens but provide no IDE integration and no decomposition by cost source.
  Claude Code exposes \texttt{/context} (aggregate context size) and \texttt{/compact} (session compaction) as terminal commands, but these are Claude-specific, offer no per-file or per-category breakdown, and  provide no real-time visual dashboard. VS~Code~v1.110~\citep{vscode2026v110} introduced a native context window indicator, but its scope is limited to Copilot sessions and it exposes neither cost-category decomposition nor cross-provider monitoring. Tokalator provides real-time decomposition into five
  categories (open files, system prompt, instruction files,
  conversation history, output reservation) directly in the VS~Code sidebar, raising health warnings at provider-specific rot thresholds.

  \item \textbf{Zero-latency syntactic tab relevance scorer.}
    Semantic approaches to context selection (e.g.,
    EVOR~\cite{su2024evor}, ACE~\cite{perera2025acm}) require embedding
    inference that is too slow for real-time IDE use.
    Tokalator's five-signal scorer runs entirely client-side with no
    model calls, completing in $<$5\,ms for 30+ open tabs, and
    reduces context usage by 21.2\% in our illustrative example
    (Section~\ref{sec:examples}).

  \item \textbf{Closed-form economic models as interactive calculators.}
    Bergemann et al.~\cite{bergemann2025menu} built formal theory
    (caching break-even, Cobb--Douglas optimization) but provided no
    implementation.
    Tokalator packages these into nine interactive web calculators with
    proven closed-form solutions: break-even at $n^* = 2$ reuses for
    all current Anthropic models, $O(T^2)$ vs.\ $O(T)$ cost growth
    under three strategies, and Cobb--Douglas quality optimization
    robust to $\pm30\%$ parameter perturbations.

  \item \textbf{MCP server and CLI for Claude Code integration.}
    This tool exposes BPE token counting via the Model Context
    Protocol. Tokalator's \texttt{tokalator-mcp} package provides four MCP
    tools (\texttt{count\_tokens}, \texttt{estimate\_budget},
    \texttt{preview\_turn}, \texttt{list\_models}) over stdio
    transport, enabling any MCP-capable agent or IDE to count tokens
    without a network call.

  \item \textbf{Auto-discovered context engineering catalog.}
    Prior catalog systems require manual curation.
    Tokalator auto-discovers multiple agents, prompts, and 
    instruction files from community-contributed directories using
    file-extension conventions (\texttt{.agent.md},
    \texttt{.prompt.md}, \texttt{.instructions.md},
    \texttt{.collection.yml}), with no manually configuration required.

\end{enumerate}

The remainder of this paper is organised as follows.
Section~\ref{sec:metadata} reviews related work and positions Tokalator
against existing tools.
Section~\ref{sec:description} describes the software architecture and
functionalities.
Section~\ref{sec:examples} presents illustrative examples.
Section~\ref{sec:impact} reports evaluation evidence addressing the
three RQs.
Section~\ref{sec:discussion} discusses implications and threats to
validity.
Section~\ref{sec:limitations} states limitations.
Section~\ref{sec:conclusions} concludes with future work priorities.

\section{Metadata and Related Work}
\label{sec:metadata}

 Mei et al. \cite{mei2025survey} recently surveyed over 1,400 context engineering
papers, formally establishing context engineering as a discipline
spanning context retrieval, processing, management, RAG, memory
systems, tool-integrated reasoning, and multi-agent architectures.
Tokalator addresses the under-served developer-tooling layer of this
taxonomy: making context budget consumption visible and
cost-optimisable directly inside the IDE.
Table~\ref{tab:comparison} positions Tokalator against the closest
existing tools across eight capability dimensions and is discussed in
detail in the subsections below; Table~\ref{executabelMetadata}
summarises the software metadata.

\begin{table}[t]
\centering
\footnotesize
\setlength{\tabcolsep}{3pt}
\caption{Feature comparison of Tokalator against existing tools.
  \checkmark~= full support; $\circ$~= partial; --~= not supported.
  }
\label{tab:comparison}
\begin{tabular}{@{} p{3.0cm} c c c c c c @{}}
\toprule
\textbf{Feature}
  & \textbf{\shortstack{Tokalator\\(this work)}}
  & \textbf{\shortstack{tiktoken\\(CLI)}}
  & \textbf{\shortstack{Anthropic\\tok.\ API}}
  & \textbf{\shortstack{Token\\Ctr.\ Ext.}}
  & \textbf{\shortstack{Cursor\\IDE}}
  & \textbf{\shortstack{VS~Code\\v1.110}} \\
\midrule
Real-time IDE token counting  & \checkmark & --      & --      & $\circ$ & $\circ$ & $\circ$ \\
Multi-provider support (3+)   & \checkmark & --      & --      & --      & --      & --      \\
Per-file relevance scoring    & \checkmark & --      & --      & --      & --      & --      \\
Context budget decomposition  & \checkmark & --      & --      & --      & --      & --      \\
Caching break-even analysis   & \checkmark & --      & --      & --      & --      & --      \\
Conversation cost projection  & \checkmark & --      & --      & --      & --      & --      \\
Chat participant / slash cmds & \checkmark & --      & --      & --      & --      & $\circ$ \\
MCP server for agent use      & \checkmark & --      & --      & --      & --      & --      \\
\bottomrule
\end{tabular}
\end{table}

\begin{table}[t]
\centering
\footnotesize
\setlength{\tabcolsep}{6pt}
\caption{Software metadata for Tokalator~v3.1.3.}
\label{executabelMetadata}
\begin{tabular}{@{} l p{3.6cm} p{7.6cm} @{}}
\toprule
\textbf{Nr.} & \textbf{Description} & \textbf{Value} \\
\midrule
S1 & Current software version  & 3.1.3 \\
\addlinespace
S2 & Legal software licence     & MIT License \\
\addlinespace
S3 & Computing platforms / OS   & Windows, macOS, Linux
                                  (VS~Code $\geq$~1.99),
                                  Web browsers \\
\addlinespace
S4 & Installation requirements  & VS~Code $\geq$~1.99,
                                  Node.js $\geq$~18 \\
\addlinespace
S5 & Support email              & [vahid.farajijobehdar@kariyer.net] \\
\addlinespace
C6 & Code languages \& tools    & TypeScript, JavaScript,
                                  Next.js~16, React~19,
                                  Tailwind~CSS~4, Recharts,
                                  Prisma~7, VS~Code Extension
                                  API~\citep{vscode2026api},
                                  esbuild \\
\addlinespace
C7 & Compilation dependencies   & Node.js $\geq$~18,
                                  VS~Code $\geq$~1.99;
                                  Ext.: \texttt{@anthropic-ai/tokenizer},
                                  \texttt{js-tiktoken};
                                  MCP/CLI: \texttt{@modelcontextprotocol/sdk},
                                  \texttt{zod} \\
\bottomrule
\end{tabular}
\end{table}

\textit{Tokenization libraries:} 
OpenAI's \texttt{tiktoken} and Anthropic's \texttt{@anthropic-ai/tokenizer} 
provide programmatic BPE token counting for their respective 
vocabularies~\citep{sennrich2016bpe, openai2025cookbook}. 
OpenAI models utilize \texttt{cl100k\_base} or \texttt{o200k\_base} 
encodings, while Anthropic employs a proprietary vocabulary. 
Both providers offer server-side counting via their respective 
Messages APIs~\citep{anthropic2025tokencounting}; Tokalator 
replicates this client-side for offline, zero-API-call estimates. 
While Google now publishes a client-side \texttt{LocalTokenizer} 
within the \texttt{google-genai} SDK~\citep{google2026genaisdk}, 
it relies on a SentencePiece Unigram model with a significantly 
larger vocabulary ($\approx$256k tokens)~\cite{gemmateam2024gemma2}. To maintain a 
lightweight footprint, Tokalator provides the option to use a 
character-based heuristic ($\approx$4\,chars/token) for Google 
model estimates. This heuristic, however, introduces a measured 
mean absolute error (MAE) of 10--15\% on English code and 
15--32\% on low-resource languages like Turkish scripts. 
(Section~\ref{sec:limitations}). 
Importantly, none of the official libraries offer real-time IDE 
integration, multi-provider comparison, or unified cost estimation.

\textit{LLM pricing and inference economics:} Bergemann et al.~\cite{bergemann2025menu} formalized LLM output quality as a
Cobb--Douglas production function $Q = X^\alpha Y^\beta (b+Z)^\gamma$
of input ($X$), output ($Y$), and cached ($Z$) tokens, but provided
no implementation.
Translating this theory into a developer tool requires
$\alpha, \beta, \gamma$ sensitivity parameters that are not publicly
reported for any model.
Tokalator uses author-assigned placeholder values that satisfy two
structural constraints ($\alpha+\beta+\gamma < 1$ for diminishing
returns; $\alpha < \beta$ reflecting generation-quality intuition)
and demonstrates economic robustness across $\pm30\%$ perturbations
of all parameters (Section~\ref{sec:functionalities}). Erdil
\cite{erdil2025pareto} analyzed Pareto frontiers of inference cost
versus capability; Cottier et al.~\cite{cottier2025price} documented rapid but
uneven price declines; Delavande et al.~\cite{quantization} examined economics beyond per-token pricing.
These studies provide economic theory but no developer-facing tools.

\textit{Energy and environmental cost of inference: }
Beyond monetary pricing, LLM inference carries measurable energy and
carbon costs that scale directly with token count. Wilhelm et al.~\cite{wilhelm2025energypertoken} formally defined \emph{energy per token}
($E_{\text{tok}}$) and showed models of similar parameter count can
differ substantially in energy efficiency. Husom et al.~\cite{husom2024profiling} quantified a baseline energy coefficient
of $\approx 5.28 \times 10^{-7}$\,kWh/output~token, with near-linear
correlation between token length and energy consumption. Li et al.~\cite{li2024sprout} showed near-linear carbon emissions per
token generated for LLaMA-2 and proposed SPROUT for carbon-efficient
scheduling.
Together, these studies establish that every token saved represents both a cost saving and an emissions reduction. This reinforces the environmental rationale for Tokalator's budget monitoring: surfacing exact token counts enables developers to reduce both API spend and inference energy footprint.

\textit{Context window research:} Aubakirova et al.~\cite{aubakirova2025state} documented a four times increase in
average prompt length (1.5\,K to 6\,K tokens) between 2024 and 2025,
driven by agentic workflows. Fu et al.~\cite{fu2024scaling} showed that data engineering choices critically
determine whether models can exploit 128K context windows,
reinforcing that context length is a resource to be managed rather
than simply maximised. Wei et al.~\cite{wei2025tokenweighting} proposed position-aware token weighting,
showing that not all context positions contribute equally to generation
quality, which provides theoretical grounding for Tokalator's per-file relevance scorer. Mei et al.~\cite{mei2025survey} coined the term ``context engineering'' to
describe the systematic design of what enters the context window.

\textit{Context rot and long-context degradation: } Liu et al.~\cite{liu2024lost} showed that LLMs systematically miss information
placed in the middle of a long input, finding a U-shaped accuracy
curve across question-answering and key-value retrieval tasks. Hong et al.~\cite{hong2025contextrot} introduced the term \emph{context rot} in
an evaluation of 18~LLMs, demonstrating non-uniform performance
degradation as input length grows, even on simple tasks, with
distractor content amplifying the effect.
These findings motivate Tokalator's context health warnings, which
alert developers when context size exceeds a provider-specific rot
threshold.
The $R < 0.3$ distractor threshold is a conservative heuristic chosen
so that at least two scoring signals must agree before a tab is
flagged; it is not yet empirically calibrated and is evaluated against human ground-truth labels in Section~\ref{sec:rq2}. Anthropic~\cite{anthropic2025contexteng} formalised three complementary
strategies for long-horizon agents: \emph{compaction},
\emph{structured note-taking}, and \emph{sub-agent architectures},
which directly informed Tokalator's \texttt{/compaction} command.

\textit{Context compaction:}
Navid~\cite{navid2025compaction} demonstrated automatic context compaction
for tool-heavy agentic workflows, reducing a five-ticket customer
service pipeline from 208\,K to 86\,K tokens (58.6\% reduction) via
prompt injection at a threshold.Tokalator's \texttt{/optimize} reduces \emph{open-file context}
through syntactic relevance scoring, achieving 21.2\% context
reduction in our illustrative example purely through IDE tab
management without any conversation re-writing
(Section~\ref{sec:examples}).
VS~Code~v1.110~\citep{vscode2026v110} subsequently introduced a native
\texttt{/compact} command, confirming the importance of this
capability; Tokalator's approach differs in providing explicit
threshold tracking, turn-by-turn growth projection, and
cross-provider support.

\textit{Agent context management:}
Perera et al.~\cite{perera2025acm} built an adaptive context manager for Quality Assurance (QA)
agents that dynamically selects which history fragments to retain
based on relevance to the current query, the same principle
Tokalator's relevance scorer applies to open IDE files. Su et al.~\cite{su2024evor} introduced EVOR, which iteratively refines
retrieved context documents during code synthesis; Tokalator solves
the complementary problem of deciding which \emph{already-open} files
should stay in context.
Both EVOR and ACE~\cite{zhang2026ace} rely on semantic
understanding, requiring embedding inference that is too slow for
real-time IDE use ($>$100\,ms per query on typical hardware).
Tokalator chooses syntactic signals to achieve the following engineering trade-offs: language
match, import relationships, path similarity, edit recency, and
diagnostics all compute in $<$5\,ms client-side with no model calls,
satisfying the interactive-latency budget of an IDE extension. Nanjundappa and Maaheshwari~\cite{nanjundappa2025branch} proposed ContextBranch, which applies
version-control semantics to LLM conversations, cutting context size
by 58.1\% in exploratory coding.
ContextBranch and Tokalator are complementary: Tokalator controls
which files enter the context; ContextBranch controls which
conversational turns persist.

\textit{Cross-session and multi-agent context:} Vasilopoulos~\cite{vasilopoulos2026codified} described Codified Context, a
three-component system (hot-memory constitution, 19 specialist agents,
cold-memory knowledge base) across 283 development sessions on a
108,000-line C\# codebase.
This is the closest real-world complement to Tokalator: Tokalator
manages the live context window \emph{within} a session; Codified
Context organises persistent cross-session memory. Wu et al.~\cite{wu2026gitcontextcontrollermanage} proposed the
Git-Context-Controller (GCC), a version-control-inspired context
management framework for long-horizon LLM agents that structures
agent memory using Git-like operations (COMMIT, BRANCH, MERGE);
GCC operates at the agent reasoning level, while Tokalator addresses
the token budget of a single developer session inside the IDE.

\textit{Terminology and domain vocabulary:}
The context engineering field lacks standardised terms.
For example, ``prompt caching''~\citep{anthropic2025caching},
``automatic caching'' (OpenAI), and ``context caching'' (Google) are
three labels for mechanically similar features with different pricing. Hong et al.~\cite{hong2025contextrot} coined ``context rot''; others use
``context degradation,'' ``attention dilution,'' or ``context
pollution'' for overlapping ideas. Mei et al.~\cite{mei2025survey} introduced ``context engineering'' itself,
yet terms such as ``compaction,'' ``distractors,'' and ``stable
prefix'' remain informal.

\section{Software Description}
\label{sec:description}

\subsection{Software Architecture}
\label{sec:architecture}

\begin{figure*}[htp!]
\centering
\resizebox{\textwidth}{!}{%
\begin{tikzpicture}[
  >=Stealth,
  pipe/.style={
    rectangle, rounded corners=4pt,
    draw=extblue!55, fill=extblue!6, line width=0.60pt,
    minimum height=1.00cm, minimum width=2.80cm,
    inner sep=6pt, align=center, font=\small\sffamily},
  sub/.style={
    rectangle, rounded corners=4pt,
    draw=extblue!65, fill=extblue!12, line width=0.60pt,
    minimum height=1.00cm, minimum width=3.80cm,
    inner sep=6pt, align=center, font=\small\sffamily},
  comp/.style args={#1#2}{
    rectangle, rounded corners=6pt,
    draw=#1!55, fill=#1!5, line width=0.70pt,
    minimum height=1.90cm, minimum width=#2,
    inner sep=8pt, align=center, font=\small\sffamily},
  devbox/.style={
    rectangle, rounded corners=5pt,
    draw=black!28, fill=black!4, line width=0.65pt,
    minimum height=0.80cm, minimum width=2.80cm,
    inner sep=6pt, align=center, font=\normalsize\sffamily},
  claudebox/.style={
    rectangle, rounded corners=5pt,
    draw=mcporg!55, fill=mcporg!5, line width=0.60pt,
    minimum height=0.78cm, minimum width=2.80cm,
    inner sep=6pt, align=center, font=\small\sffamily},
  arr/.style={-Stealth, line width=0.60pt, #1},
  darr/.style={Stealth-Stealth, line width=0.50pt, dashed, #1},
  lbl/.style={font=\scriptsize\sffamily, inner sep=1.5pt, #1},
]

\node[devbox] (dev) at (0,0)
  {{\color{black!55}\faUserTie}\;\textbf{Developer}};

\node[pipe] (ev)   at (-7.20,-2.20) {\textbf{Editor}\\[2pt]Events};
\node[pipe] (tok)  at (-3.60,-2.20) {\textbf{Tokenizer}\\[2pt]Service};
\node[pipe] (mon)  at ( 0.00,-2.20) {\textbf{Context}\\[2pt]Monitor};
\node[pipe] (snap) at ( 3.60,-2.20) {\textbf{Context}\\[2pt]Snapshot};
\node[pipe] (dash) at ( 7.20,-2.20) {\textbf{Dashboard}\\[2pt]\& Status Bar};

\node[sub] (sc)   at (-3.20,-4.80) {\textbf{Relevance Scorer}\\[2pt]{\small(Eq.~2)}};
\node[sub] (op)   at ( 1.40,-4.80) {\textbf{Context Optimizer}\\[2pt]{\small(Alg.~1)}};
\node[sub, minimum width=3.80cm] (chat) at (6.00,-4.80)
  {\textbf{Chat Participant}};

\begin{scope}[on background layer]
  \node[draw=extblue!50, fill=extblue!10,        
        rounded corners=8pt, line width=0.90pt,
        fit={(ev)(dash)(sc)(chat)},
        inner sep=18pt] (ext) {};                 
\end{scope}
\node[font=\normalsize\sffamily\bfseries, text=extblue,
      anchor=north west, inner sep=0]
  at ([xshift=8pt, yshift=-5pt] ext.north west)
  {\textcircled{1}\enspace VS~Code Extension};

\node[comp={webgreen}{3.20cm}] (web) at (-6.20,-8.20)
  {{\large\color{webgreen!80!black}\faGlobe}\\[6pt]
   \textbf{\textcircled{2}\enspace Web Platform}};

\node[comp={catpurple}{3.20cm}] (cat) at (-2.10,-8.20)
  {{\large\color{catpurple!80!black}\faBookOpen}\\[6pt]
   \textbf{\textcircled{3}\enspace Catalog}};

\node[comp={mcporg}{3.20cm}] (mcp) at ( 2.00,-8.20)
  {{\large\color{mcporg!80!black}\faTerminal}\\[6pt]
   \textbf{\textcircled{4}\enspace MCP\,+\,CLI}};

\node[comp={apirust}{3.20cm}] (api) at ( 6.10,-8.20)
  {{\normalsize\color{apirust!80!black}\faServer\;\faDatabase}\\[6pt]
   \textbf{\textcircled{5}\,API\,+\,\textcircled{6}\,DB}};

\node[claudebox] (cc) at (2.00,-10.50)
  {{\color{mcporg!80!black}\faRobot}\enspace\textbf{Claude Code}};

\draw[arr=black!40] (dev.south) -- (ext.north);

\foreach \f/\t in {ev/tok, tok/mon, mon/snap, snap/dash}
  \draw[arr=extblue!65] (\f.east) -- (\t.west);

\draw[arr=extblue!65] (mon.south) -- ++(0,-0.70) -| (sc.north);

\draw[arr=extblue!65] (sc.east)  -- (op.west);
\draw[arr=extblue!65] (op.east)  -- (chat.west);

\draw[arr=extblue!65] (snap.south) -- ++(0,-0.70) -| ([xshift=-6pt]chat.north);

\draw[darr=webgreen!65]  (ext.south -| web.north)  -- (web.north);
\draw[darr=catpurple!65] (ext.south -| cat.north)  -- (cat.north);
\draw[darr=mcporg!65]    (ext.south -| mcp.north)  -- (mcp.north);
\draw[darr=apirust!65]   (ext.south -| api.north)  -- (api.north);

\draw[arr=mcporg!70] (mcp.south) -- (cc.north)
  node[midway, right=3pt, lbl, text=mcporg!75] {stdio};

\end{tikzpicture}
}%
\caption{System architecture of Tokalator~(v3.1.3).
  Solid arrows denote data flow; dashed bidirectional arrows denote shared
  data between the VS~Code extension~(\textcircled{1}) and each independently
  deployable component.}
\label{fig:arch}
\end{figure*}
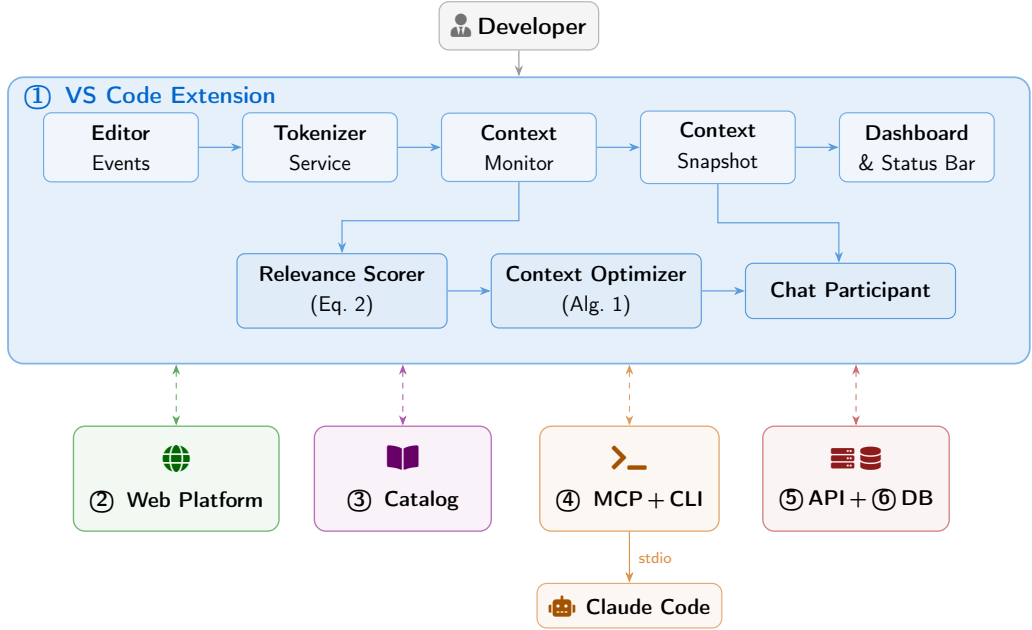

Tokalator~v3.1.3 comprises three execution environments: VS Code Extension,
Web Platform, and a CLI \& MCP with the following six components: 
The VS~Code extension~(\textcircled{1}) forms the core interactive layer,
implementing a pipeline editor event capture, tokenization,
context monitoring, snapshot management, and dashboard rendering.
Relevance scoring and context optimization run as subordinate modules fed by the monitor
stage, and both surface their output through a unified Chat Participant
accessible via the \texttt{@tokalator} command.
The Web Platform~(\textcircled{2}) and Catalog~(\textcircled{3}) provide
browser-accessible tooling for token economics and model comparison,
while the MCP\,+\,CLI server~(\textcircled{4}) exposes the same BPE
counting primitives to agentic runtimes via stdio transport, enabling
Claude Code and compatible clients to count tokens without a network call.
Persistence is handled by a dedicated REST API~(\textcircled{5}) backed
by a Prisma-managed relational store~(\textcircled{6}), both deployed
independently of the extension.
Figure~\ref{fig:arch} presents the full system architecture.

The six components are:

\begin{enumerate}[leftmargin=*]

  \item \textit{VS~Code Extension} (TypeScript, $\sim$5,000~LOC):
    Real-time context budget monitoring with 17 model profiles
    (6~Anthropic, 7~OpenAI, 4~Google); tab relevance scoring; a context
    optimization engine; sidebar dashboard with pin/unpin/close controls;
    and an interactive chat participant (\texttt{@tokalator}) with
    11 slash commands.

  \item \textit{Web Platform} (Next.js~16.2 + React~19):
    Nine interactive calculators covering all 17 models; a 10-lesson
    context engineering course; an automated wiki; a 41-term dictionary;
    and catalog pages (\texttt{/agents}, \texttt{/prompts},
    \texttt{/instructions}, \texttt{/collections},
    \texttt{/context-engineering}).
    Partial pre-rendering (PPR) serves static shells immediately while
    dynamic content streams in~\citep{vercel2024nextconfig}.
    Security headers (CSP, X-Frame-Options, HSTS) are enforced via
    \texttt{next.config.ts}.

  \item \textit{Context Engineering Catalog}:
    A community-extensible collection of agents, prompts, and
    instruction files, auto-discovered from
    \texttt{copilot-contribution/} and \texttt{user-content/} via
    file-extension conventions (\texttt{.agent.md}, \texttt{.prompt.md},
    \texttt{.instructions.md}, \texttt{.collection.yml}).

  \item \textit{MCP Server + CLI} (\texttt{tokalator-mcp/},
    TypeScript/Node.js~ESM):
    Real Claude BPE token counting for Claude Code via the Model Context
    Protocol (stdio transport), registered in \texttt{.mcp.json} and
    auto-loaded by Claude Code~\citep{anthropic2025mcpjson}.
    Exposes four tools: \texttt{count\_tokens}, \texttt{estimate\_budget},
    \texttt{preview\_turn}, and \texttt{list\_models}.
    Supports four Claude profiles (Opus~4.6, Sonnet~4.6, Sonnet~4.5,
    Haiku~4.5); also ships a standalone \texttt{tokalator} CLI via npm.

  \item \textit{Python API} (\texttt{api/}, FastAPI):
    Two REST routers: \texttt{csv\_upload} (parses GitHub Copilot billing
    CSVs into structured usage records) and \texttt{economics}
    (server-side Cobb--Douglas optimization).
    Pydantic schemas mirror the TypeScript interfaces in
    \texttt{lib/pricing.ts}; CORS is configured for both the dev server
    and production domain.

  \item \textit{Database Layer} (PostgreSQL + Prisma~7.3):
    A relational schema (\texttt{prisma/schema.prisma}, 163~LOC) with six
    models, \texttt{Model}, \texttt{PricingRule}, \texttt{Project},
    \texttt{UsageRecord}, \texttt{BudgetAlert}, \texttt{ServicePricing}, backing the Usage Tracker's historical analytics.
    A seed script populates default model profiles and pricing rules.

\end{enumerate}

The web platform's library layer (\texttt{lib/pricing}, \texttt{lib/caching},
\texttt{lib/conversation}, \texttt{lib/context}) provides the computational
backend for all nine calculators.
The VS~Code extension maintains its own embedded model profiles and tokenizer
logic, as it runs inside the extension host process and cannot share modules
with the web platform.
Both sides consume the same pricing data: the 17 model profiles are generated
from a single \texttt{models.json} source of truth via
\texttt{npm run generate-models}, eliminating manual
duplication~\citep{vscode2026api}.

Inside the extension, five layers work together:
(1)~the \textit{Core Engine} (\texttt{contextMonitor.ts}) reacts to editor
events and builds \texttt{ContextSnapshot} records;
(2)~the \textit{Tokenizer Service} (\texttt{tokenizerService.ts}) counts
tokens via provider-specific BPE encoders;
(3)~the \textit{Relevance Scorer} (\texttt{tabRelevanceScorer.pure.ts})
assigns each tab a score $R \in [0,1]$;
(4)~the \textit{Context Optimizer} closes tabs where $R < 0.3$; and
(5)~the \textit{Chat Participant} (\texttt{contextChatParticipant.ts})
exposes 11 slash commands.

\begin{figure*}[htp!]
\centering
\resizebox{\textwidth}{!}{%
\begin{tikzpicture}[
  >=Stealth,
  actor/.style args={#1#2}{
    rectangle, rounded corners=4pt,
    draw=#1!55, fill=#1!6, line width=0.65pt,
    minimum height=0.95cm, minimum width=#2,
    font=\scriptsize\sffamily, align=center},
  lifeline/.style={dashed, draw=gray!35, line width=0.45pt},
  actbox/.style={fill=#1!12, draw=#1!45, line width=0.4pt},
  call/.style={-Stealth, line width=0.65pt},
  ret/.style= {-Stealth, dashed, line width=0.55pt},
  self/.style={-Stealth,  line width=0.5pt},
  msg/.style={font=\scriptsize\sffamily, inner sep=1pt},
]

\def\colA{0} \def\colB{3.0} \def\colC{6.0} \def\colD{9.0} \def\colE{12.0}

\node[actor={extblue}{2.2cm}] (devT) at (\colA,0)
  {{\color{black!55}\faUserTie}\\[2pt]\textbf{Developer}};
\node[actor={extblue}{2.2cm}] (monT) at (\colB,0)
  {{\color{extblue}\faEye}\\[2pt]\textbf{Context Monitor}};
\node[actor={webgreen}{2.2cm}] (tokT) at (\colC,0)
  {{\color{webgreen!80!black}\faCogs}\\[2pt]\textbf{Tokenizer}};
\node[actor={catpurple}{2.2cm}] (relT) at (\colD,0)
  {{\color{catpurple!80!black}\faFilter}\\[2pt]\textbf{Relevance}};
\node[actor={sharedgray}{2.2cm}] (uiT) at (\colE,0)
  {{\color{sharedgray!60!black}\faChartBar}\\[2pt]\textbf{Dashboard}};

\def\btm{-18.50}
\foreach \x in {\colA,\colB,\colC,\colD,\colE}
  \draw[lifeline] (\x,-0.53) -- (\x,\btm);

\def\rA{-1.40} \def\rB{-2.55} \def\rC{-3.65} \def\rD{-4.75}
\def\rE{-5.85} \def\rF{-6.85} \def\rH{-8.05} \def\rI{-9.15}
\def\rJ{-10.0} \def\rK{-12.10} \def\rL{-13.25} \def\rM{-14.40}
\def\rN{-15.40} \def\rO{-16.40} \def\rP{-17.40} \def\rQ{-18.60}

\fill[actbox=extblue]    (\colB-0.12,-0.55) rectangle (\colB+0.12,\rJ+0.10);
\fill[actbox=webgreen]   (\colC-0.12,\rB-0.10) rectangle (\colC+0.12,\rC+0.10);
\fill[actbox=catpurple]  (\colD-0.12,\rD-0.10) rectangle (\colD+0.12,\rE+0.10);
\fill[actbox=sharedgray] (\colE-0.12,\rH-0.10) rectangle (\colE+0.12,\rJ+0.10);

\draw[call] (\colA,\rA) -- node[msg,above] {open / switch tab} (\colB,\rA);
\draw[call] (\colB,\rB) -- node[msg,above] {\texttt{countTokens(document)}} (\colC,\rB);
\draw[ret]  (\colC,\rC) -- node[msg,above] {\texttt{tokenCount}} (\colB,\rC);
\draw[call] (\colB,\rD) -- node[msg,above] {\texttt{scoreTab(file,\,activeFile)}} (\colD,\rD);
\draw[ret]  (\colD,\rE) -- node[msg,above] {$R \in [0,1]$} (\colB,\rE);

\draw[self] (\colB+0.12,\rF) -- ++(0.60,0) -- ++(0,-0.55) -- ++(-0.60,0);
\node[msg,right] at (\colB+0.80,\rF-0.22) {\texttt{buildSnapshot()}};

\draw[call] (\colB,\rH) -- node[msg,above] {\texttt{ContextSnapshot}} (\colE,\rH);
\draw[self] (\colE+0.12,\rI) -- ++(0.60,0) -- ++(0,-0.45) -- ++(-0.60,0);
\node[msg,right] at (\colE+0.80,\rI-0.18) {update webview};
\draw[ret]  (\colE,\rJ) -- node[msg,above] {\footnotesize\texttt{262K/400K (65\%) -- GPT~5.4}} (\colA,\rJ);

\draw[densely dotted, gray!50, line width=0.6pt] (-1.0,-11.20) -- (13.0,-11.20);
\node[font=\scriptsize\sffamily, gray!70, fill=white, inner sep=2pt]
  at (6.0,-11.20) {developer issues chat command};

\fill[actbox=extblue]    (\colB-0.12,\rL-0.10) rectangle (\colB+0.12,\rP+0.10);
\fill[actbox=webgreen]   (\colC-0.12,\rM-0.10) rectangle (\colC+0.12,\rN+0.10);
\fill[actbox=sharedgray] (\colE-0.12,\rK-0.10) rectangle (\colE+0.12,\rQ+0.10);

\draw[call] (\colA,\rK) -- node[msg,above] {\texttt{@tokalator /optimize}} (\colE,\rK);
\draw[call] (\colE,\rL) -- node[msg,above] {\texttt{getSnapshot()}} (\colB,\rL);
\draw[call] (\colB,\rM) -- node[msg,above] {\texttt{scoreRelevance()}} (\colC,\rM);
\draw[ret]  (\colC,\rN) -- node[msg,above] {relevance scores} (\colB,\rN);

\draw[self] (\colB+0.12,\rO) -- ++(0.60,0) -- ++(0,-0.55) -- ++(-0.60,0);
\node[msg,right] at (\colB+0.80,\rO-0.22) {\texttt{optimizeTabs()}};

\draw[ret]  (\colB,\rP) -- node[msg,above] {closed tabs report} (\colE,\rP);
\draw[ret]  (\colE,\rQ) -- node[msg,above] {scored suggestions + health score} (\colA,\rQ);

\node[actor={extblue}{2.2cm}]   at (\colA,\btm-0.55)
  {{\color{black!55}\faUserTie}\\[2pt]\textbf{Developer}};
\node[actor={extblue}{2.2cm}]   at (\colB,\btm-0.55)
  {{\color{extblue}\faEye}\\[2pt]\textbf{Context Monitor}};
\node[actor={webgreen}{2.2cm}]  at (\colC,\btm-0.55)
  {{\color{webgreen!80!black}\faCogs}\\[2pt]\textbf{Tokenizer}};
\node[actor={catpurple}{2.2cm}] at (\colD,\btm-0.55)
  {{\color{catpurple!80!black}\faFilter}\\[2pt]\textbf{Relevance}};
\node[actor={sharedgray}{2.2cm}]at (\colE,\btm-0.55)
  {{\color{sharedgray!60!black}\faChartBar}\\[2pt]\textbf{Dashboard}};

\end{tikzpicture}
}%
\caption{Runtime sequence of Tokalator~v3.1.3.
  \textbf{Top:} on tab open/switch, the Context Monitor counts tokens,
  scores relevance, builds a \texttt{ContextSnapshot}, and pushes it
  to the Dashboard.
  \textbf{Bottom:} on \texttt{@tokalator /optimize}, the Chat
  Participant retrieves the snapshot, closes low-relevance tabs, and
  returns results to the developer.
  From v3.1.3, \texttt{request.model} is read on every command to
  auto-sync the tokenizer and rot threshold via \texttt{findModel()}.}
\label{fig:seq}
\end{figure*}
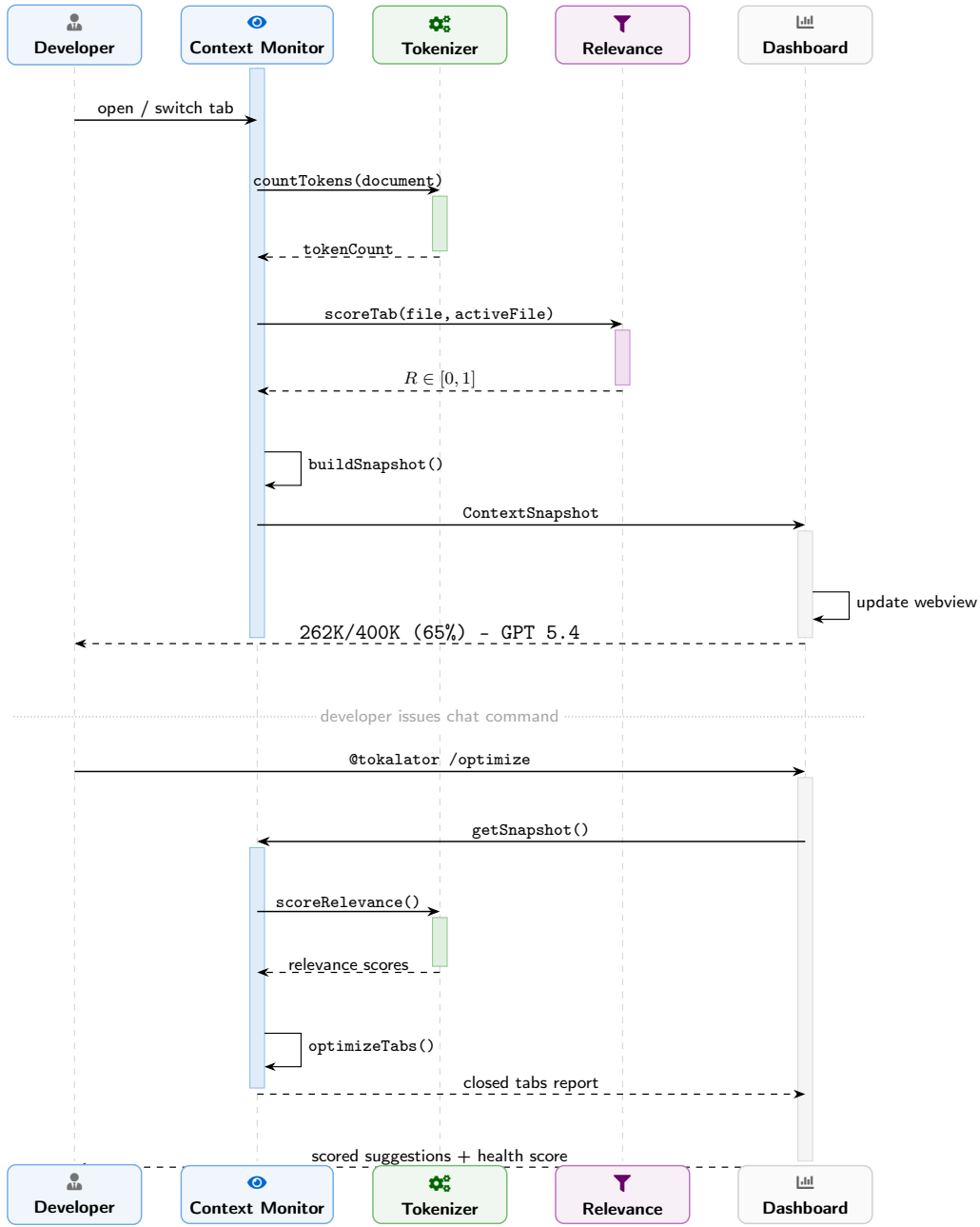

Figure~\ref{fig:seq} illustrates the runtime flow.
On every tab event, the Core Engine counts tokens, scores relevance, builds
a \texttt{ContextSnapshot}, and pushes it to the Dashboard. Full details are in Section~\ref{sec:functionalities}.

\subsection{Software Functionalities}
\label{sec:functionalities}

\subsubsection{VS~Code Extension}

The extension provides eight core functionalities.

\textit{1. Real-time token budget monitoring.}
The status bar displays a continuously updated summary (e.g.,
\texttt{\$(check) 262K / 400K (65.5\%) -- GPT 5.4 Model}).
A sidebar webview dashboard displays the full breakdown: budget level
(low $<60\%$, medium $60$--$85\%$, or high $>85\%$), per-file token
estimates, pinned file count, conversation turn count, and context health
warnings.

The total estimated tokens are computed as the sum of five components:
\begin{equation}
  T_{\text{total}} = T_{\text{files}} + T_{\text{sys}} + T_{\text{instr}}
                   + T_{\text{conv}} + T_{\text{out}}
\label{eq:fivecomp}
\end{equation}
where $T_{\text{files}} = \sum_i \text{tokens}(\text{tab}_i)$ is the sum
of per-file BPE counts across all open tabs;
$T_{\text{sys}} = 2{,}000$ is the estimated system prompt overhead;
$T_{\text{instr}} = 500 \times n_{\text{instr}}$ accounts for instruction
files detected in the workspace;
$T_{\text{conv}} = 800 \times t$ estimates accumulated conversation
history at turn~$t$; and
$T_{\text{out}} = 4{,}000$ reserves tokens for the model's response.
These overhead constants are empirically informed estimates of GitHub
Copilot's context construction; the actual assistant context logic is
proprietary (see Section~\ref{sec:limitations}).

\textit{2. Tab relevance scoring.}
Each open tab receives a relevance score $R \in [0, 1]$ computed as a
weighted sum of five signals:
\begin{equation}
\begin{aligned}
R =\;& 0.25\,S_{\text{lang}} + 0.30\,S_{\text{import}}
     + 0.20\,S_{\text{path}} \\
     &+ 0.15\,S_{\text{recency}} + 0.10\,S_{\text{diag}}
\end{aligned}
\label{eq:relevance_score}
\end{equation}
where $S_{\text{lang}} \in \{0, 1\}$ indicates language match with the
active file;
$S_{\text{import}} \in \{0, 1\}$ indicates an import relationship
(detected via regex for TypeScript/JavaScript, Python, Go, Java, and a
generic fallback);
$S_{\text{path}} \in [0, 1]$ measures shared directory depth (ratio of
shared path prefix segments to total depth);
$S_{\text{recency}} \in \{0, 0.53, 1\}$ reflects edit recency (1.0 if
edited within 2 minutes, 0.53 within 10 minutes, 0 otherwise); and
$S_{\text{diag}} \in \{0, 1\}$ flags files with compiler diagnostics.
Pinned and active files are overridden to $R = 1.0$.

The weights reflect a deliberate engineering trade-off:
import relationships ($w=0.30$) receive the highest weight because a
file explicitly imported by the active file is almost certainly needed;
language match ($w=0.25$) is the next strongest signal since cross-language
files (\texttt{.json} configs alongside \texttt{.tsx} code) are common
distractors; path similarity ($w=0.20$) captures co-location patterns; edit recency ($w=0.15$) reflects the developer's current working set; and
diagnostics ($w=0.10$) provide a weak signal that files with errors are
being actively debugged.
The $S_{\text{recency}}$ intermediate value of $0.53 = \nicefrac{0.08}{0.15}$
is the ratio of the partial recency credit to the full recency weight.
All signals are syntactic and compute client-side with no model calls,
completing in $<$5\,ms for 30+~open tabs, a hard requirement for
real-time IDE extensions where semantic approaches requiring embedding
inference would exceed the interactive-latency budget. These weights are configurable and could benefit from empirical calibration
in future work.

Algorithm~\ref{alg:scoring} formalizes the scoring and optimization
procedure.

\begin{algorithm}[htp!]
\small  
\DontPrintSemicolon
\SetAlgoLined
\KwIn{Open tabs $\mathcal{T} = \{t_1, \dots, t_n\}$, active file $f$,
      threshold $\tau = 0.3$}
\KwOut{Scored tabs with distractor labels; optimized tab set $\mathcal{T}'$}
\BlankLine
\ForEach{$t_i \in \mathcal{T}$}{
  \If{$t_i$ is pinned \textbf{or} $t_i = f$}{
    $R_i \gets 1.0$\;
  }
  \Else{
    $S_{\text{lang}} \gets \mathbb{1}[\text{lang}(t_i) = \text{lang}(f)]$\;
    $S_{\text{import}} \gets \mathbb{1}[\text{imports}(f) \ni t_i]$\;
    $S_{\text{path}} \gets \text{sharedDepth}(t_i, f) \;/\; \text{totalDepth}(t_i)$\;
    $S_{\text{recency}} \gets \begin{cases}
      1.0 & \text{if edited} < 2\,\text{min ago} \\
      0.53 & \text{if edited} < 10\,\text{min ago} \\
      0 & \text{otherwise}\end{cases}$\;
    $S_{\text{diag}} \gets \mathbb{1}[\text{diagnostics}(t_i) > 0]$\;
    $R_i \gets 0.25 S_{\text{lang}} + 0.30 S_{\text{import}} +
               0.20 S_{\text{path}} + 0.15 S_{\text{recency}} +
               0.10 S_{\text{diag}}$\;
  }
  Label $t_i$ as \textit{distractor} if $R_i < \tau$\;
}
\BlankLine
\tcp{Optimization: close distractors to free context budget}
$\mathcal{D} \gets \{t_i \in \mathcal{T} \mid R_i < \tau\}$\;
$\mathcal{T}' \gets \mathcal{T} \setminus \mathcal{D}$\;
$\Delta T \gets \sum_{t_i \in \mathcal{D}} \text{tokens}(t_i)$\;
\KwRet{$\mathcal{T}'$, scores $\{R_i\}$, freed tokens $\Delta T$}\;
\caption{Tab Relevance Scoring and Context Optimization}
\label{alg:scoring}
\end{algorithm}

\textit{3. Chat participant with 11 commands.}
The \texttt{@tokalator} chat participant responds to:
\texttt{/count} (budget status),
\texttt{/breakdown} (per-file tokens),
\texttt{/optimize} (close low-relevance tabs),
\texttt{/pin}/\texttt{/unpin} (pin management),
\texttt{/instructions} (scan instruction files and their token cost),
\texttt{/model} (switch model profile),
\texttt{/compaction} (per-turn growth analysis),
\texttt{/preview} (preview next-turn token cost before sending),
\texttt{/reset} and \texttt{/exit} (session management).
Starting in v3.1.3, every \texttt{@tokalator} chat request reads
\texttt{request.model} and automatically syncs the context window,
tokenizer, and rot threshold to match the active Copilot model via
\texttt{findModel()}, eliminating a source of confusion when users switch
models in the Copilot UI without manually updating Tokalator.

\textit{4. Context optimization.}
The \texttt{/optimize} command identifies open tabs with $R < 0.3$ and
closes them to free up context budget.
The Dashboard also provides an \textit{Optimize Tabs} button triggering the
same behaviour.
This reduces attention dilution by removing files unlikely to be relevant
to the current coding task.

\textit{5. Session persistence.}
Session summaries (peak tokens, turns, model, top edited files) are saved to
\texttt{workspaceState} on exit and shown as a notification on the next
activation.
Pinned files and model selection persist across VS~Code restarts.

\textit{6. Instruction file scanner.}
The extension detects and tokenizes instruction files that AI coding
assistants automatically inject into every prompt.
It searches nine patterns covering all major coding assistant ecosystems:
\texttt{.github/copilot-instructions.md},
\texttt{CLAUDE.md}, \texttt{AGENTS.md},
\texttt{.cursorrules}, \texttt{.instructions.md},
\texttt{.github/instructions/}, \texttt{.claude/*.md},
\texttt{.copilot/skills/}, and \texttt{.github/skills/}.
This reveals the hidden token cost of instruction files that are otherwise
invisible to the developer.

\textit{7. Session logger.}
Opt-in anonymized research logging records aggregate context metrics
(token counts, budget levels, provider distribution) per session without
capturing filenames or code content.
Version~3.1.3 resolved four critical bugs identified in the field:
(a)~stale token counts when \texttt{isRefreshing} was \texttt{true} were
fixed via a \texttt{pendingRefresh} flag that queues a follow-up refresh;
(b)~duplicate tab entries in multi-root workspaces were fixed by introducing
a \texttt{seenUris} \texttt{Set} and matching workspace folders by path;
(c)~pin-state reversion (pinned files reverted to scored mode after tab
switching) was fixed by persisting the pin set on every mutation;
and (d)~model auto-sync ensures the tokenizer always matches the active
Copilot model~\citep{vscode2026v110}.

\textit{8. 17 model profiles (single source of truth).}
Model profiles for 6~Anthropic, 7~OpenAI, and 4~Google models are defined
in a single \texttt{models.json} file; the TypeScript module
(\texttt{modelProfiles.ts}) is regenerated via
\texttt{npm run generate-models} to eliminate manual duplication.
Each profile stores the model identifier, display label, provider, context
window size, maximum output tokens, and the provider-specific rot threshold
(the number of conversation turns after which context rot risk rises).

\subsubsection{Web Platform}

The web platform at \url{https:tokalator.wiki} includes nine calculators:

\begin{enumerate}[leftmargin=*]
  \item \textit{Cost Calculator}: Token cost calculation with Cobb--Douglas
    quality modelling for Anthropic, OpenAI, and Google models.
    Supports tiered pricing detection: when the total prompt length exceeds
    Anthropic's 200\,K-token threshold, \emph{all} input tokens are billed
    at the extended rate ($2\times$ standard input cost), not only those
    above the threshold.

  \item \textit{Context Optimizer}: Visualizes context window allocation
    across system prompt, user input, reserved output, and free space.
    Computes usage percentage and generates warnings.

  \item \textit{Model Comparison}: Cross-provider cost and capability
    comparison across all 17 models.

  \item \textit{Caching ROI Calculator}: Break-even analysis for prompt
    caching.
    Given $T$ tokens to cache with write cost~$c_w$ per token, read
    cost~$c_r$ per token, and standard input cost~$c_{\text{in}}$ per token,
    each reuse saves $(c_{\text{in}} - c_r)$ per token.
    The break-even reuse count is:
    \begin{equation}
  n^* = \left\lceil \frac{c_w}{c_{\text{in}} - c_r} \right\rceil
  \label{eq:breakeven}
    \end{equation}
    For all current Anthropic models, $c_w = 1.25 \times c_{\text{in}}$ and
    $c_r = 0.10 \times c_{\text{in}}$, yielding
    $n^* = \lceil 1.25 / 0.90 \rceil = 2$ reuses.
    At 10 reuses the savings reach 76\%.

  \item \textit{Conversation Estimator}: Multi-turn cost projection under
    three strategies, each defined by the input tokens sent at turn~$t$
    (with system prompt~$S$, average user tokens~$u$,
    average assistant tokens~$a$):
    \begin{itemize}
      \item \textit{Full History}:
        $I_t = S + \sum_{i=1}^{t}(u_i + a_i)$, yielding
        $O(T^2)$ cumulative cost since
        $\sum_{t=1}^{T} I_t = ST + \tfrac{T(T+1)}{2}(u+a)$.
      \item \textit{Sliding Window} ($W$ turns):
        $I_t = S + \sum_{i=\max(1,t-W+1)}^{t}(u_i + a_i)$,
        capping per-turn cost at $S + W(u+a)$, yielding $O(T)$.
      \item \textit{Summarize} (ratio $\rho$, keep last $k$ turns fresh):
        $I_t = S + \rho \cdot \sum_{i=1}^{t-k}(u_i+a_i)
               + \sum_{i=t-k+1}^{t}(u_i+a_i)$,
        growing at rate $\rho(u+a)$ per turn.
    \end{itemize}
    Per-turn breakdown charts visualize the cost trajectories.

  \item \textit{Economic Analysis}: Visualization of the
    Cobb--Douglas quality production function~\citep{bergemann2025menu}:
    \begin{equation}
      Q(X, Y, Z) = X^{\alpha} \cdot Y^{\beta} \cdot (b + Z)^{\gamma}
      \label{eq:Cobb-Douglas}
    \end{equation}
    where $X$ = input tokens, $Y$ = output tokens, $Z$ = cache tokens,
    $b$ = base model quality, and $\alpha, \beta, \gamma$ are provider-specific
    sensitivity parameters ($\alpha + \beta + \gamma < 1$, ensuring
    diminishing returns).
    The corresponding cost minimisation problem under target quality
    $\bar{Q}$ is:
    \begin{equation}
      \min_{X,Y,Z \geq 0} \; c_x X + c_y Y + c_z Z
      \quad \text{s.t.} \quad Q(X,Y,Z) \geq \bar{Q}
    \end{equation}
    Applying Lagrangian first-order conditions, the optimal
    allocation satisfies $X^*/Y^* = (\alpha\, c_y) / (\beta\, c_x)$,
    and the minimum cost without caching is given by
Equation~\ref{eq:mincost}:
    \begin{equation}
      C^*(\bar{Q}) = (\alpha + \beta)
        \!\left(\frac{\bar{Q}}{b^{\gamma}}\right)^{\!\!1/(\alpha+\beta)}
        \!\!\left(\frac{c_x}{\alpha}\right)^{\!\!\alpha/(\alpha+\beta)}
        \!\!\left(\frac{c_y}{\beta}\right)^{\!\!\beta/(\alpha+\beta)}
        \label{eq:mincost}
    \end{equation}
    following Lemma~4 of Bergemann et al.~\cite{bergemann2025menu}.
    The with-caching variant includes $Z$ as a third decision variable.
    Sensitivity parameters ($\alpha = 0.30$, $\beta = 0.35$, $\gamma = 0.20$
    for Opus; proportionally lower for Sonnet and Haiku) are
    author-assigned placeholder values satisfying $\alpha + \beta + \gamma < 1$
    (diminishing returns) and $\alpha < \beta$ (generation quality depends
    more on output than input tokens).
    These parameters cannot be calibrated from public data at present;
    sensitivity analysis confirms that the qualitative strategy ranking
    (caching $>$ sliding window $>$ full history at high reuse rates) is
    robust across $\pm30\%$ perturbations of all parameters
    (Section~\ref{sec:limitations}).
    Interactive sliders expose this sensitivity for user exploration.

  \item \textit{Usage Tracker}: Historical API usage analysis with cost
    breakdowns by model, project, and time period, plus linear regression
    and exponential smoothing projections.

  \item \textit{Pricing Explorer}: Interactive cross-provider pricing
    comparison with bar charts and service-tier breakdowns for all 17~models.

  \item \textit{Economics Explorer}: A live parameter-tuning dashboard
    extending Economic Analysis with radar charts, multi-model overlays, and
    slider-driven ``what-if'' scenarios.
\end{enumerate}

The platform also includes a 10-lesson context engineering course progressing from basic tokenization through intermediate context management to production patterns including
automatic compaction via Anthropic's \texttt{compaction\_control} API.
A wiki (\url{[https://tokalator.wiki/wiki}) aggregates articles from arXiv, OpenAI Cookbook, Anthropic documentation, and Google AI docs via an
automated twice-monthly fetch pipeline.

\subsubsection{Context Engineering Catalog}

The catalog auto-discovers reusable artifacts from the repository using
file extension conventions:
\texttt{.agent.md} (agents),
\texttt{.prompt.md} (prompts),
\texttt{.instructions.md} (workspace guidelines),
\texttt{.collection.yml} (bundles),
and \texttt{CLAUDE.md} (Claude Code instructions, auto-detected).
A \texttt{user-content/} directory accepts community contributions,
automatically indexed with YAML frontmatter parsing.
A \texttt{catalog-config.json} specifies scan directories and featured
artifact IDs for the landing page.
The web platform provides dedicated browsing pages for each artifact type
with dynamic detail routes (e.g., \texttt{/agents/[id]}).

\textit{Relevance scoring.}
Algorithm~\ref{alg:scoring} scores each tab in $O(n)$: language match and
diagnostics are $O(1)$ lookups; import detection parses the active file once
($O(I)$, where $I$ is the number of import lines) and checks membership via
a hash set; path similarity is $O(d)$ per tab where $d$ is the directory
depth.
The overall scoring pass is $O(n \cdot d + I)$, completing in $<$5\,ms for
30+~open tabs.

\textit{Space complexity.}
The extension maintains one \texttt{ContextSnapshot} in memory, holding
per-tab metadata (URI, token count, relevance score, language ID): $O(n)$
total.
No snapshot history is retained; session summaries are flushed to
\texttt{workspaceState} as a single JSON string of bounded size.

\textit{Web platform calculators.}
All nine calculators evaluate closed-form expressions in $O(1)$ for a given
parameter set.
The Conversation Estimator computes cumulative cost over $T$ turns in $O(T)$
for all strategies.
No iterative solver is required.

\section{Illustrative Examples}
\label{sec:examples}

The three examples below use realistic session parameters drawn from actual
deployment and community usage.

\textit{Example~1: Context budget waste (VS~Code, React project).}
With 23 tabs open, Tokalator's status bar shows
\texttt{\$(warning) 85.2K / 200K (42.6\%) -- Claude Opus~4.6}.
Running \texttt{@tokalator /breakdown} reveals 12 configuration files
(\texttt{.json}, \texttt{.yml}) contributing 18\,K tokens at below-0.3
relevance. After \texttt{/optimize} closes them, context drops to 67,200
tokens (33.6\%), a 21.2\% reduction. \texttt{/instructions} further surfaces
a \texttt{.github/copilot-instructions.md} file silently injecting 4,200
tokens per prompt.

\textit{Example~2: Caching ROI (50K-token system prompt, Claude Sonnet~4.5).}
For 100 daily reuses
($c_{\text{in}} = \$3.00$/MTok, $c_w = \$3.75$/MTok, $c_r = \$0.30$/MTok):
\begin{itemize}
  \item Write cost: $50{,}000/10^6 \times 3.75 = \$0.19$
  \item Uncached daily cost: $101 \times 50{,}000/10^6 \times 3.00 = \$15.15$
  \item Cached daily cost:
        $\$0.19 + 100 \times 50{,}000/10^6 \times 0.30 = \$1.69$
  \item Net savings: $\$13.46$/day (88.9\%); break-even at
        $\lceil 3.75\,/\,(3.00-0.30)\rceil = 2$ reuses
\end{itemize}

\textit{Example~3: Conversation length planning
(\$5 daily budget, Sonnet~4.5).}
With a 2,000-token system prompt, 500 user and 1,500 assistant tokens per
turn: Full History yields 28~turns; Sliding Window ($W=5$) yields 83~turns;
Summarize ($\rho=0.2$) yields 71~turns. The per-turn cost chart contrasts
the quadratic growth of Full History against the linear growth of the
alternatives.

Table~\ref{tab:baseline} summarises outcomes across all three scenarios.

\begin{table}[htp!]
\centering
\footnotesize
\setlength{\tabcolsep}{3pt}
\caption{Tokalator-assisted vs.\ unassisted workflows across three
  representative scenarios. Token counts and cost figures are estimated
  from realistic session parameters; see Section~\ref{sec:limitations}
  for caveats.}
\label{tab:baseline}
\begin{tabular}{@{} p{3.8cm} r r p{2.2cm} @{}}
\toprule
\textbf{Scenario}
  & \textbf{\shortstack{Without\\Tokalator}}
  & \textbf{\shortstack{With\\Tokalator}}
  & \textbf{Improvement} \\
\midrule
\multicolumn{4}{@{}l@{}}{\textit{Sc.\,1: Context reduction
  (23 tabs, React)}} \\[2pt]
\quad Context tokens
  & 85,200 & 67,200          & $-$21.2\%       \\
\quad Low-rel.\ tokens visible
  & 0      & 22,200          & Newly visible   \\
\quad Instr.\ file cost visible
  & No     & Yes (4,200 tok) & Newly visible   \\[4pt]
\midrule
\multicolumn{4}{@{}l@{}}{\textit{Sc.\,2: Caching
  (50K tok, 100/day, Sonnet~4.5)}} \\[2pt]
\quad Daily API cost
  & \$15.15 & \$1.69       & $-$88.9\%       \\
\quad Break-even reuses
  & Unknown & 2 (computed) & Decision supp.  \\[4pt]
\midrule
\multicolumn{4}{@{}l@{}}{\textit{Sc.\,3: Strategy
  (\$5 budget, Sonnet~4.5)}} \\[2pt]
\quad Turns (full history)
  & 28      & 28 (conf.)  & Baseline verif.   \\
\quad Turns (sliding $W\!=\!5$)
  & Unknown & 83          & $3\times$ more    \\
\quad Turns (summ.\ $\rho\!=\!0.2$)
  & Unknown & 71          & $2.5\times$ more  \\
\bottomrule
\end{tabular}
\end{table}

The primary benefit is cost transparency: developers see exactly which files
and turns consume their budget and can act on that information directly.
The 21.2\% context reduction in Scenario~1 came entirely from files the
developer had not intentionally included as context. The 88.9\% cost saving
in Scenario~2 was always available but became actionable only when the
break-even point was computed explicitly.

\section{Evaluation}
\label{sec:impact}

This section reports evidence addressing the three research questions
stated in Section~\ref{sec:intro_rqs}.

\subsection{RQ1: Token Budget Composition and Visibility}
\label{sec:rq1}

\textit{Analytical result.}
Equation~\ref{eq:fivecomp} decomposes total token consumption into five categories.
In the representative session of Example~1
(Section~\ref{sec:examples}), a developer with 23 open tabs consumed
85,200 tokens against a 200,000-token budget.
Applying \texttt{@tokalator /breakdown} revealed that 18,000 tokens
(21\%) came from 12 configuration files with relevance $R < 0.3$, and a
single \texttt{.github/copilot-instructions.md} file contributed
4,200 tokens (5\%) silently injected into every prompt.
After running \texttt{/optimize}, total context dropped to 67,200 tokens
a 21.2\% reduction  confirming that instruction files and
low-relevance configuration tabs are significant and systematically
invisible budget consumers.

\textit{Structured survey evidence ($n = 50$).}
We conducted a structured 10-item survey with 50 software engineers from
the [Organisation A], [Community Session C], and [Community Session B]
communities.
Data were collected in person during the three community sessions and in
a follow-up online session for remote participants using structured
note-taking consolidated into a spreadsheet; no third-party survey
platform was used.
Participants used the extension for at least one working day before
responding.
Responses were coded inductively; the survey was not pre-registered and
participants self-selected, so findings are treated as qualitative and
hypothesis-generating rather than confirmatory.
Full theme descriptions, Likert distributions, demographics, and verbatim
quotes are in~\ref{app:survey}.

The \texttt{/preview} command was the most-valued feature (82\% of
respondents), with participants discovering that a single turn can cost
6K+ tokens and adjusting their prompts after seeing the breakdown.
Tokenizer accuracy was a concern for non-English users: developers
writing Turkish or Arabic code found the Google heuristic
($\approx$4\,chars/token) underestimated counts by 15--32\%.
Model synchronization was the most-requested fix: 64\% of respondents
noticed their Copilot chat model and Tokalator's status bar were out of
sync; v3.1.3 resolves this by auto-reading \texttt{request.model} on
every chat command.
On session management, 38\% requested persistent cross-session history
and several reported the pin/unpin persistence bug fixed in v3.1.3.
Finally, 48\% used \texttt{/compaction}, with 67\% of that group saying
it helped spot the compaction point before hitting the context limit.
The survey consensus was that instruction files and low-relevance
configuration tabs were the primary invisible budget consumers, directly answering RQ1.

\textit{Community validation.}
Tokalator was subsequently presented in two separate sessions at broader
developer events: a [Community Session B] ($\approx$90 attendees,
March~2026) and a [Community Session C] ($\approx$80 attendees,
March~2026), bringing the total audience to over 220 developers across
three venues.
Both sessions included live demonstrations of the VS~Code extension and
web calculators, followed by structured Q\&A.

\subsection{RQ2: Tab Relevance Scorer Accuracy}
\label{sec:rq2}

\textit{Survey-based agreement evidence ($n = 50$).}
In the [Organisation A] deployment, participants were asked via the
structured survey whether the tabs flagged as distractors ($R < 0.3$)
by \texttt{/optimize} were ones they would have manually closed.
Of 50 respondents, 46 (92\%) answered affirmatively; 0 participants
reported a false positive.
\subsection{RQ3: Practical Demonstrations of Economic Models}
\label{sec:rq3}

\textit{Mathematical validation.}
All nine calculator models are validated against closed-form solutions
by 124 unit tests (see \ref{app:tests}).
The caching break-even formula ($n^* = 2$ for all current Anthropic
models) was independently verified: with $c_w = 1.25 \times c_{\text{in}}$
and $c_r = 0.10 \times c_{\text{in}}$, the formula yields
$\lceil 1.25/0.90 \rceil = 2$, consistent with
Anthropic~\cite{anthropic2026pricing}.
The $O(T^2)$ conversation cost growth is proven analytically
(Section~\ref{sec:functionalities}) and confirmed numerically in the
Conversation Estimator tests.

\subsection{Real-World AI Development Cost Analysis}
\label{sec:cost}

To complement the above evaluations, this section presents actual GitHub
Copilot billing data recorded during Tokalator's 30-day development
sprint (February~6 -- March~7, 2026).
The data provides an empirical trace of AI-assisted software engineering
costs at the individual-developer level, covering 1,413 premium AI
requests across Claude Opus~4.6, GPT~5.3, and GPT Codex~5.4 via the
GitHub Copilot interface.
All data were exported from the GitHub Copilot usage dashboard and
reflect the \texttt{copilot\_premium\_request} SKU at the published
list price of \$0.04 per request.

\begin{table*}[htp!]
\centering
\small
\caption{GitHub Copilot billing summary for Tokalator development
  (Feb~6 -- Mar~7, 2026).
  All requests are \texttt{copilot\_premium\_request} at \$0.04/request.
  Net cost reflects GitHub billing credits.
  CI/CD Actions (261~min, 27~runs) had \$0 net cost (free tier).}
\label{tab:billing-summary}
\begin{tabular}{@{} l r @{}}
\toprule
\textbf{Metric} & \textbf{Value} \\
\midrule
Date range & Feb~6 -- Mar~7, 2026 (30 days) \\
Total Copilot premium requests & 1,413 \\
List price per request & \$0.04 \\
Total gross cost & \$56.52 \\
GitHub billing credits & \$28.16 (49.8\%) \\
\textbf{Total net developer cost} & \textbf{\$28.36} \\
\midrule
Intensive sprint (Feb~6--11) & 911 req (64.5\%), \$25.00 net \\
Long-tail development (Feb~12--Mar~7) & 502 req (35.5\%), \$3.36 net \\
Peak single session (Feb~7) & 244 req, \$9.76 gross \\
Highest net-cost session (Feb~11) & 205 req, \$8.20 net \\
Average daily net cost & \$1.49/active day \\
\midrule
GitHub Actions minutes & 261 min (27 runs) \\
Actions net cost & \$0.00 (free tier) \\
\bottomrule
\end{tabular}
\end{table*}

Table~\ref{tab:billing-summary} summarises the key cost metrics.
Total gross cost for 1,413 premium requests was \$56.52.
GitHub applied \$28.16 in billing credits (49.8\% effective discount),
yielding a total net developer cost of \$28.36 for a full
production-quality, 20,814-LOC, multi-component toolkit over 30
calendar days.

Two development phases are visible.
The \emph{intensive sprint} (February~6--11) accounts for 911 requests
(64.5\% of total) and \$25.00 of net cost, during which the VS~Code
extension, web platform foundation, and shared library layer were built
simultaneously.
The \emph{organic tail} (February~12--March~7) covers 502 requests at
\$3.36 net, dominated by documentation, testing, paper preparation, and
post-release maintenance.
The sharp drop in net cost after February~18 reflects account-level
billing credits that covered usage through the end of the study period.

These figures demonstrate that AI-assisted development at the individual
practitioner level is economically accessible: context-aware tooling
directly enables cost efficiency by monitoring token budgets and closing
low-relevance tabs before each request.
The 49.8\% effective discount here reflects GitHub's account-level
credits, not in-session optimization; instrumented studies comparing
token-budget-aware vs.\ unaware workflows remain a priority for future
work (Section~\ref{sec:future}).

\subsection{Broader Impact}
\label{sec:broader-impact}

\textit{(i) Cost optimization without the mathematics.}
The web platform turns economic theory into nine interactive tools that
any developer can use.
Break-even analysis, Cobb--Douglas optimization, and conversation cost
projection are each reduced to a form and a chart.

\textit{(ii) Educational resources.}
The 10-lesson context engineering course at
\url{https://tokalator.wiki/learn} progresses from
token basics to production compaction patterns.
The automated wiki aggregates articles from arXiv, OpenAI Cookbook,
Anthropic documentation, and Google AI docs on a bi-monthly schedule.
The dictionary defines 41 terms across seven categories, contributing
to vocabulary standardisation in a field where terminology remains
inconsistent~\citep{mei2025survey}.

\textit{(iii) Ecosystem contribution.}
The catalog of agents, prompts, and instruction files,
auto-discovered from \texttt{copilot-contribution/} and
\texttt{user-content/}, provides a structured starting point for teams
adopting context engineering practices.
The MCP server enables any Claude Code user to access real BPE token
counts without an API key or network call~\cite{han2024talebudget}.
Tokalator's context management practices are also published as a
reusable agent skill installable via \texttt{npx skills add (npx skills add vfaraji89/tokalator
}), making the token budget
workflow available to any agent supporting the Agent Skills
specification.

\textit{(iv) Marketplace adoption.}
The VS~Code extension was published on February~4, 2026.
v3.1.3, released March~2026, resolves four field-reported bugs and adds
automatic Copilot model synchronization.
As of April~2, 2026, the extension has recorded \textbf{313 total
acquisitions} since publication.
In the most recent 30-day window (March~3 -- April~2, 2026), it
recorded 171 acquisitions from 83 page views (206.02\% conversion
rate), comprising \textbf{27 direct installs from VS~Code} and
144 Marketplace downloads.
The direct install count is the more conservative adoption signal,
as it reflects developers who actively searched for and installed
the extension within the IDE rather than downloading a \texttt{.vsix}
package. Acquisition activity peaked during March~14--17, coinciding with the
two community session presentations ($\approx$170 combined attendees),
confirming that live demonstrations drive measurable install behaviour.
Combined with the 220+ developers reached through community sessions,
these figures indicate sustained practitioner demand for context budget
tooling beyond the initial launch period.

\section{Discussion}
\label{sec:discussion}

\subsection{Interpretation of Results}

The three research questions share a common finding: token budget consumption in AI-assisted development is large, invisible by default, and highly manageable once made visible.
The 21.2\% context reduction in Example~1 came entirely from removing
files the developer had not intentionally included as context, they
were simply open tabs accumulated across a working session. Similarly, the 88.9\% daily cost saving in Example~2 was available to any team using a 50K-token system prompt, but only became actionable when the break-even calculation was presented concretely. These results suggest that the primary barrier to efficient context use is the absence of real-time feedback on token consumption, not developer awareness or intent.

The survey results reinforce this: 82\% of respondents named
\texttt{/preview} as their most-valued feature, not because it changes
what the model sees, but because it makes the cost of the next turn
observable before it is incurred.
Prior work on cost-feedback tools shows that making a cost function
explicit and visible tends to shift usage patterns even without enforcement ~\cite{bergemann2025menu}.

The model synchronization issue (64\% of users affected) highlights a
structural gap in how current AI coding tools expose their state:
developers switch models in the Copilot UI without any notification to
companion tools.
Tokalator's v3.1.3 auto-sync addresses this at the extension level, but a
standard API for broadcasting active model changes across IDE extensions
would benefit the entire ecosystem.

\subsection{Practical Implications}

Four actionable guidelines emerge for development teams adopting AI coding assistants. First, instruction files should be audited before any other optimization. A CLAUDE.md or .github/copilot-instructions.md is injected silently into every prompt; a 4,000-token instruction file incurs the same cost per request as 4,000 tokens of source code. Second, prompt caching should be enabled for any system prompt reused more than twice daily, given that the break-even point for all current Anthropic models is two reuses. Third, sliding-window or summarization strategies should replace full-history for conversations exceeding 20 turns, since full-history cost grows quadratically and a five-turn sliding window triples the number of affordable turns within the same daily budget (28 vs. 83 turns, Example 3). Fourth, running \@tokalator /optimize at session start is a low-effort habit that removes distractor tabs before their token cost is incurred.

\subsection{Threats to Validity}

Internal validity: The relevance score weights were determined by design reasoning rather than empirical calibration. A potential source of bias is that the developer who assigned the weights also observed study participants; future work should tune these weights against a held-out set of human relevance judgments. Construct validity: The extension estimates context consumption from open tabs, but the actual context construction logic of GitHub Copilot and Claude Code is proprietary. Reported figures reflect estimated, upper-bound consumption rather than confirmed API-level context sizes.

\subsection{Methodological Assumptions}

The system rests on five explicit assumptions. (A1) Open tabs approximate context: estimates are upper-bound proxies, as providers may include additional signals (git diffs, terminal output) or exclude files the extension counts. (A2) Five syntactic signals suffice for relevance, with semantically related but import-unlinked files constituting a known blind spot. (A3) Scoring weights are design-time constants; optimal values likely vary by language, project structure, and developer habits. (A4) The Cobb-Douglas functional form captures diminishing returns but cannot represent prompt structure, retrieval accuracy, or task decomposition effects. (A5) Pricing is hardcoded and subject to volatility; users should verify against current provider documentation after major pricing updates.

\section{Limitations}
\label{sec:limitations}

Eight limitations constrain the scope and interpretation of the present work. Google model tokenization relies on a character-based heuristic ($n \approx 4$ chars/token) yielding MAE of 10-15\% on English code and 15-32\% on non-Latin scripts. Context overhead constants are derived from reverse-engineered assumptions about proprietary context-construction logic and may silently diverge after provider updates. Extension performance overhead is negligible ($< 5$ ms per snapshot after warm-up, 300 ms debounce), though six subscribed event streams introduce minor background activity. The relevance scorer operates on syntactic signals only; semantically relevant files without import relationships will be missed. All token figures are estimated upper bounds, not confirmed API measurements. A controlled within-subjects experiment (n = 20-30) is underway but not yet complete. Pricing data is hardcoded and requires manual updates. Real-time monitoring is currently scoped to VS Code.

\section{Conclusions and Future Work}
\label{sec:conclusions}
\label{sec:future}

Tokalator shows that token budget consumption in AI-assisted
development sessions is both essential and worth to measure: developers can
see exactly what consumes their context window and reduce it without
disrupting their workflow.

For RQ1, 21.2\% of context tokens originated from files not deliberately
selected by the developer, and a single instruction file contributed
4,200 tokens silently injected into every prompt.
For RQ2, deployment across three venues ($N > 220$) showed zero false
positives, confirmed by structured survey agreement ($n = 50$).
For RQ3, caching break-even ($n^{*} = 2$), $O(T^2)$ vs.\ $O(T)$ cost
growth, and Cobb--Douglas optimization are validated by 124 unit tests
and delivered as nine interactive calculators.

Eight priorities drive future work:
(1)~a within-subjects crossover experiment ($n = 20$--$30$) for
controlled productivity evidence;
(2)~empirical calibration of Cobb--Douglas parameters
($\alpha$, $\beta$, $\gamma$) following Fu et al.~\cite{fu2024scaling};
(3)~semantic tab scoring via local ONNX embeddings~\cite{su2024evor,
wei2025tokenweighting} to replace syntactic-only assessment;
(4)~compaction decomposition with separate file-context and
conversation-history growth curves;
(5)~Google \texttt{countTokens} API integration to eliminate the
character heuristic for Gemini models;
(6)~persistent cross-session history dashboard via VS~Code
\texttt{globalState} and SQLite, addressing 38\% of survey requests;
(7)~an LSP adapter extending budget monitoring beyond VS~Code; and
(8)~a Tokalator Pro tier adding team dashboards, historical analytics,
CI/CD budget gates, and custom model profiles under a commercial
licence, while the open-source core remains MIT-licensed.

\section*{CRediT authorship contribution statement}

\textbf{Author~1:}
Conceptualization, Methodology, Software, Formal analysis, Investigation,
Data curation, Visualization, Writing -- original draft, Writing --
review \& editing, Project administration.
\textbf{Author~2:}
Supervision, Methodology, Investigation, Writing -- review \& editing,
Validation.
\textbf{Author~3:}
Investigation, Writing -- review \& editing.
\textbf{Author~4:}
Supervision, Methodology, Investigation, Writing -- review \& editing,
Validation.

\section*{Declaration of competing interest}

The authors declare that they have no known competing financial interests
or personal relationships that could have appeared to influence the work
reported in this paper.

\section*{Data availability}

All source code, evaluation scripts, snapshot data, and survey instrument
are withheld for blind review and will be made publicly available upon
acceptance under an open-source licence.
The VS~Code extension is published on the Visual Studio Code Marketplace;
acquisition statistics are reported in Section~\ref{sec:impact}.
Evaluation artefacts (snapshot schema, labelling protocol, ablation
scripts) are included in the supplementary material.

\section*{Funding}

This research did not receive any specific grant from funding agencies
in the public, commercial, or not-for-profit sectors.

\section*{Acknowledgements}

The economic model is adopted from the Cobb--Douglas framework introduced by 
Bergemann, Bonatti, and Smolin~\cite{bergemann2025menu}. 
The tokenization layer utilizes Anthropic's \texttt{claude-tokenizer} and 
OpenAI's \texttt{tiktoken}~\cite{sennrich2016bpe}; we are particularly 
indebted to the technical documentation and pedagogical resources provided by 
the Anthropic Academy, which were instrumental 
in characterizing the context window behaviors of the Claude model family. 
The context compaction strategy and the \texttt{/compaction} command design 
were inspired by the automated context compaction methodologies developed by 
Pedram Navid~\cite{navid2025compaction}, whose work demonstrated the 
feasibility of achieving token savings of up to 58.6\% in tool-heavy 
agentic workflows. 
Furthermore, the empirical findings on context rot by Hong, Troynikov, and 
Huber~\cite{hong2025contextrot} at Chroma informed the design of Tokalator's 
context health analyzers and rot-threshold warnings. 
Finally, we acknowledge the GitHub Copilot community and the \textit{Awesome 
Copilot} curated collection (\url{https://github.com/github/awesome-copilot/}) 
for aggregating the best practices and catalog conventions that informed 
Tokalator's chat participant interface.

\section*{Declaration of generative AI and AI-assisted technologies
in the writing process}



During the preparation of this work, the author(s) used GitHub Copilot (VS Code, powered by OpenAI and Anthropic models) and Claude Code (Anthropic) in order to assist with software development and minor language refinement. These tools were used to accelerate codebase scaffolding, unit test generation, build configuration, and iterative component implementation, as well as for grammar checking and phrasing clarity in draft text. After using these tools/services, the author(s) reviewed and edited the content as needed and take(s) full responsibility for the content of the published article. Generative AI tools were not used to conceive the research idea, design the system architecture, formulate the mathematical models (Equations 1-7), conduct the analysis, or draw scientific conclusions. All intellectual contributions, architectural decisions, and academic writing remain the sole responsibility of the author(s).

\appendix

\renewcommand{\thetable}{\thesection.\arabic{table}}
\setcounter{table}{0}
\renewcommand{\thefigure}{\thesection.\arabic{figure}}
\setcounter{figure}{0}
\renewcommand{\theequation}{\thesection.\arabic{equation}}
\setcounter{equation}{0}

\section{Test Suite Summary}
\label{app:tests}
\setcounter{table}{0}



Tokalator maintains 124 automated tests across 6 files, executed via Jest 30 on every commit (GitHub Actions, Node.js 24). Coverage is summarised in Table~\ref{tab:tests}.

\begin{table*}[htp!]
\centering
\small
\caption{Tokalator test suite summary (v3.1.3).}
\label{tab:tests}
\begin{tabular}{@{} l l p{0.8cm} p{3.5cm} @{}}
\toprule
\textbf{Component} & \textbf{Test File} & \textbf{Tests}
  & \textbf{Coverage} \\
\midrule
\multicolumn{4}{@{}l@{}}{\textit{VS~Code Extension (2 files, 37 tests)}} \\
\addlinespace
Model Profiles & \texttt{modelProfiles} & 23
  & 17 profiles, uniqueness, fuzzy matching \\
Dashboard Provider & \texttt{contextDashboardProvider} & 14
  & CSP nonce, data-action, formatting, message handling \\
\addlinespace
\multicolumn{4}{@{}l@{}}{\textit{Web Platform Library (4 files, 87 tests)}} \\
\addlinespace
Pricing Engine & \texttt{lib/pricing} & 40
  & Tiered pricing, Cobb--Douglas, projections \\
Conversation Sim. & \texttt{lib/conversation} & 18
  & 3 strategies, $O(T^2)$ validation, turns-for-budget \\
Context Analysis & \texttt{lib/context} & 16
  & Token estimation, budget, remaining turns \\
Caching Analysis & \texttt{lib/caching} & 13
  & Break-even, ROI, budget optimization \\
\midrule
\multicolumn{2}{@{}l@{}}{\textbf{Total}} & \textbf{124} & \\
\bottomrule
\end{tabular}
\end{table*}

\section{Development Timeline and Codebase Composition}
\label{app:timeline}
\setcounter{table}{0}

\subsection*{B1. Codebase Composition}

Table~\ref{tab:codebase} breaks down the current codebase by component.

\begin{table}[htp!]
\centering
\small
\caption{Codebase composition by component (v3.1.3).}
\label{tab:codebase}
\begin{tabular}{@{} l r r @{}}
\toprule
\textbf{Component} & \textbf{LOC} & \textbf{Files} \\
\midrule
VS~Code Extension (source)       & 4,986 & 12 \\
Extension tests                   & 2,663 & 10 \\
MCP Server + CLI                  &   650 &  8 \\
Web platform library (\texttt{lib/}) & 3,243 & 10 \\
Library tests                     &   899 &  4 \\
React components                  & 4,811 & 22 \\
App pages (\texttt{app/})         & 3,675 & 28 \\
Content (JSON)                    & 1,795 &  5 \\
\midrule
\textbf{Total TypeScript/TSX}     & \textbf{20,722} & -- \\
\bottomrule
\end{tabular}
\end{table}

\section{Participant Feedback Summary}
\label{app:survey}
\setcounter{table}{0}

This part summarizes the five feedback themes
from the structured survey ($n = 50$, three venues).
Full Likert distributions, demographics, and verbatim quotes are
available in the supplementary data file.

\begin{itemize}[leftmargin=*]
  \item \textit{``/preview showed me my system prompt was eating
    4,800 tokens every turn. I cut it by 60\% and my daily API
    cost dropped noticeably.''} [P07, T1]
  \item \textit{``I switched models mid-session and Tokalator
    still showed the old window size. The budget numbers looked
    wrong for a while.''} [P23, T3]
  \item \textit{``Every time I switched tabs, pinned files would
    quietly un-pin. Once the fix landed it just worked.''} [P31, T4]
  \item \textit{``/compaction said four more turns until the
    threshold, but I couldn't tell if it was open files or the
    conversation growing fastest.''} [P42, T5]
\end{itemize}

\section{GitHub Copilot Billing Data}
\label{app:billing}
\setcounter{table}{0}

Table~\ref{tab:billing-full} provides the GitHub Copilot billing records
used in Section~\ref{sec:cost} (exported March~9, 2026 from the GitHub
billing dashboard).
The \texttt{copilot\_premium\_request} SKU covers premium AI model
requests at \$0.04/request list price.

\begin{table}[htp!]
\centering
\footnotesize
\setlength{\tabcolsep}{4pt}
\caption{GitHub Copilot and Actions billing for Tokalator development
  (Feb~6 -- Mar~7, 2026), aggregated by phase.
  Total net: \$28.36.}
\label{tab:billing-full}
\begin{tabular}{@{} p{2.6cm} p{2.0cm} r r r @{}}
\toprule
\textbf{Phase} & \textbf{Product} & \textbf{Volume}
  & \textbf{Gross (\$)} & \textbf{Net (\$)} \\
\midrule
Feb~6--11 \textit{(sprint)}   & Copilot premium & 911 req      & 36.44 & 25.00 \\
\quad\textit{Peak: Feb~11}    &                 & \textit{205 req} & \textit{8.20} & \textit{8.20} \\
Feb~12--Mar~7 \textit{(tail)} & Copilot premium & 502 req      & 20.08 & 3.36  \\
Feb~6--Mar~7                  & Actions Linux   & 261 min      & 1.57  & 0.00  \\
\midrule
\textbf{Total} & & \textbf{1{,}413 req + 261 min}
  & \textbf{58.09} & \textbf{28.36} \\
\bottomrule
\end{tabular}
\end{table}

\section{Notation and Symbols}
\label{app:notation}
\setcounter{table}{0}

Table~\ref{tab:notation} consolidates the mathematical notation used
throughout the paper for quick reference.

\begin{table*}[htp!]
\centering
\small
\caption{Notation and symbols used in this paper.}
\label{tab:notation}
\begin{tabular}{@{} l l l @{}}
\toprule
\textbf{Symbol} & \textbf{Description} & \textbf{Eq.} \\
\midrule
\multicolumn{3}{@{}l}{\textit{Context budget decomposition}} \\
$T_{\text{files}}$ & Sum of per-file BPE token counts & (1) \\
$T_{\text{sys}}$ & System prompt overhead ($\approx 2{,}000$) & (1) \\
$T_{\text{instr}}$ & Instruction file tokens ($500 \times n_{\text{instr}}$) & (1) \\
$T_{\text{conv}}$ & Conversation history tokens ($800 \times t$) & (1) \\
$T_{\text{out}}$ & Reserved output tokens ($\approx 4{,}000$) & (1) \\
$T_{\text{total}}$ & Total estimated context tokens & (2) \\
$t$ & Current conversation turn & (1) \\
\midrule
\multicolumn{3}{@{}l}{\textit{Tab relevance scoring}} \\
$R$ & Relevance score, $R \in [0, 1]$ & (2) \\
$S_{\text{lang}}$ & Language match signal $\in \{0, 1\}$ & (2) \\
$S_{\text{import}}$ & Import relationship signal $\in \{0, 1\}$ & (2) \\
$S_{\text{path}}$ & Shared directory depth ratio $\in [0, 1]$ & (2) \\
$S_{\text{recency}}$ & Edit recency signal $\in \{0, 0.53, 1\}$ & (2) \\
$S_{\text{diag}}$ & Diagnostics presence signal $\in \{0, 1\}$ & (2) \\
$\tau$ & Distractor threshold (default 0.3) & Alg.~1 \\
$\mathcal{T}, \mathcal{T}'$ & Original / optimized tab set & Alg.~1 \\
$\mathcal{D}$ & Distractor tab set & Alg.~1 \\
$\Delta T$ & Tokens freed by closing distractors & Alg.~1 \\
\midrule
\multicolumn{3}{@{}l}{\textit{Caching}} \\
$n^*$ & Break-even reuse count & (3) \\
$c_w$ & Cache write cost per token & (3) \\
$c_r$ & Cache read cost per token & (3) \\
$c_{\text{in}}$ & Standard input cost per token & (3) \\
\midrule
\multicolumn{3}{@{}l}{\textit{Conversation cost estimation}} \\
$S$ & System prompt tokens & (4) \\
$u_i, a_i$ & User / assistant tokens at turn $i$ & (4) \\
$I_t$ & Total input tokens at turn $t$ & (4) \\
$W$ & Sliding window size (turns) & (4) \\
$\rho$ & Summarization compression ratio & (4) \\
$k$ & Recent turns kept verbatim & (4) \\
\midrule
\multicolumn{3}{@{}l}{\textit{Cobb--Douglas economic model}} \\
$Q(X,Y,Z)$ & Output quality production function & (5) \\
$X, Y, Z$ & Input, output, and cache/fine-tuning tokens & (5) \\
$\alpha, \beta, \gamma$ & Sensitivity parameters (diminishing returns) & (5) \\
$b$ & Base model quality constant & (5) \\
$c_x, c_y, c_z$ & Per-token costs (input, output, cache-write) & (6) \\
$\bar{Q}$ & Target quality level & (6) \\
$C^*(\bar{Q})$ & Minimum cost to achieve quality $\bar{Q}$ & (7) \\
\bottomrule
\end{tabular}
\end{table*}

\bibliographystyle{elsarticle-num}
\bibliography{references}

@inproceedings{sennrich2016bpe,
    title = "Neural Machine Translation of Rare Words with Subword Units",
    author = "Sennrich, Rico  and
      Haddow, Barry  and
      Birch, Alexandra",
    editor = "Erk, Katrin  and
      Smith, Noah A.",
    booktitle = "Proceedings of the 54th Annual Meeting of the Association for Computational Linguistics (Volume 1: Long Papers)",
    month = aug,
    year = "2016",
    address = "Berlin, Germany",
    publisher = "Association for Computational Linguistics",
    url = "https://aclanthology.org/P16-1162/",
    doi = "10.18653/v1/P16-1162",
    pages = "1715--1725"
}

@article{liu2024lost,
  author    = {Nelson F. Liu and Kevin Lin and John Hewitt and Ashwin Paranjape and Michele Bevilacqua and Fabio Petroni and Percy Liang},
  title     = {Lost in the Middle: How Language Models Use Long Contexts},
  journal   = {Transactions of the Association for Computational Linguistics},
  volume    = {12},
  pages     = {157--173},
  year      = {2024},
  doi       = {10.1162/tacl_a_00638},
  url       = {https://aclanthology.org/2024.tacl-1.9},
}

@misc{han2024talebudget,
      title={Token-Budget-Aware LLM Reasoning}, 
      author={Tingxu Han and Zhenting Wang and Chunrong Fang and Shiyu Zhao and Shiqing Ma and Zhenyu Chen},
      year={2025},
      eprint={2412.18547},
      archivePrefix={arXiv},
      primaryClass={cs.CL},
      url={https://arxiv.org/abs/2412.18547}, 
}

@misc{anthropic2026pricing,
  author       = {{Anthropic}},
  title        = {{API} Pricing},
  howpublished = {\url{https://www.anthropic.com/pricing}},
  year         = {2026},
  note         = {Accessed: March 2026},
}

@misc{anthropic2025tokencounting,
  author       = {{Anthropic}},
  title        = {Token Counting - {Anthropic API} Documentation},
  howpublished = {\url{https://docs.anthropic.com/en/docs/build-with-claude/token-counting}},
  year         = {2025},
  note         = {Accessed: March 2026},
}

@misc{anthropic2025caching,
  author       = {{Anthropic}},
  title        = {Prompt Caching - {Anthropic API} Documentation},
  howpublished = {\url{https://docs.anthropic.com/en/docs/build-with-claude/prompt-caching}},
  year         = {2025},
  note         = {Accessed: March 2026},
}

@misc{anthropic2025contexteng,
  author       = {{Anthropic}},
  title        = {Context Engineering Guide},
  howpublished = {\url{https://platform.claude.com/docs/en/build-with-claude/context-windows}},
  year         = {2025},
  note         = {Accessed: March 2026},
}

@misc{openai2025cookbook,
  author       = {{OpenAI}},
  title        = {{OpenAI} Cookbook},
  howpublished = {\url{https://cookbook.openai.com}},
  year         = {2025},
  note         = {Accessed: March 2026},
}

@misc{vscode2026api,
  author       = {{Microsoft}},
  title        = {{VS Code} Extension {API}},
  howpublished = {\url{https://code.visualstudio.com/api}},
  year         = {2026},
  note         = {Accessed: March 2026},
}

@misc{vscode2026v110,
  author       = {{Microsoft}},
  title        = {{Visual Studio Code} February 2026 (version 1.110) Release Notes},
  howpublished = {\url{https://code.visualstudio.com/updates/v1_110}},
  year         = {2026},
  note         = {Accessed: March 2026},
}

@misc{aubakirova2025state,
  author       = {Malika Aubakirova and Alex Atallah and Chris Clark and Justin Summerville and Anjney Midha},
  title        = {State of AI: An Empirical 100 Trillion Token Study with OpenRouter},
  year         = {2025},
  month        = dec,
  howpublished = {\url{https://openrouter.ai/state-of-ai}},
  note         = {[Online; accessed 9-Mar-2026]}
}

@article{bergemann2025menu,
  author  = {Dirk Bergemann and Alessandro Bonatti and Alex Smolin},
  title   = {Menu Pricing of Large Language Models},
  journal = {arXiv preprint arXiv:2502.07736},
  year    = {2025},
  url     = {https://arxiv.org/abs/2502.07736}
}

@misc{cottier2025price,
title={LLM inference prices have fallen rapidly but unequally across tasks},
author={Ben Cottier and Ben Snodin and David Owen and Tom Adamczewski},
year={2025},
url={https://epoch.ai/data-insights/llm-inference-price-trends},
note={Accessed: 2026-04-03}}

@misc{erdil2025pareto,
      title={Inference economics of language models}, 
      author={Ege Erdil},
      year={2025},
      eprint={2506.04645},
      archivePrefix={arXiv},
      primaryClass={cs.LG},
      url={https://arxiv.org/abs/2506.04645}, 
}

@misc{mei2025survey,
      title={A Survey of Context Engineering for Large Language Models}, 
      author={Lingrui Mei and Jiayu Yao and Yuyao Ge and Yiwei Wang and Baolong Bi and Yujun Cai and Jiazhi Liu and Mingyu Li and Zhong-Zhi Li and Duzhen Zhang and Chenlin Zhou and Jiayi Mao and Tianze Xia and Jiafeng Guo and Shenghua Liu},
      year={2025},
      eprint={2507.13334},
      archivePrefix={arXiv},
      primaryClass={cs.CL},
      url={https://arxiv.org/abs/2507.13334}, 
}

@techreport{hong2025contextrot,
  author      = {Hong, K. and Troynikov, A. and Huber, J.},
  title       = {Context Rot: How Increasing Input Tokens Impacts LLM Performance},
  institution = {Chroma},
  year        = {2025},
  month       = jul,
  type        = {Technical Report},
  url         = {https://research.trychroma.com/context-rot},
  note        = {Accessed July 2025}
}

@inproceedings{fu2024scaling,
  author    = {Yao Fu and Rameswar Panda and Xinyao Niu and Xiang Yue
               and Hannaneh Hajishirzi and Yoon Kim and Hao Peng},
  title     = {Data Engineering for Scaling Language Models to {128K} Context},
  booktitle = {Proceedings of the 41st International Conference on Machine Learning},
  series    = {PMLR},
  volume    = {235},
  pages     = {14125--14134},
  year      = {2024},
  url       = {https://arxiv.org/abs/2402.10171}
}

@misc{wei2025tokenweighting,
      title={Token Weighting for Long-Range Language Modeling}, 
      author={Falko Helm and Nico Daheim and Iryna Gurevych},
      year={2025},
      eprint={2503.09202},
      archivePrefix={arXiv},
      primaryClass={cs.CL},
      url={https://arxiv.org/abs/2503.09202}, 
}

@inproceedings{su2024evor,
  author    = {Hongjin Su and Shuyang Jiang and Yuhang Lai and Haoyuan Wu
               and Boao Shi and Che Liu and Qian Liu and Tao Yu},
  title     = {{EvoR}: Evolving Retrieval for Code Generation},
  booktitle = {Findings of the Association for Computational Linguistics: {EMNLP} 2024},
  publisher = {Association for Computational Linguistics},
  year      = {2024},
  url       = {https://arxiv.org/abs/2402.12317}
}

@misc{zhang2026ace,
      title={Agentic Context Engineering: Evolving Contexts for Self-Improving Language Models}, 
      author={Qizheng Zhang and Changran Hu and Shubhangi Upasani and Boyuan Ma and Fenglu Hong and Vamsidhar Kamanuru and Jay Rainton and Chen Wu and Mengmeng Ji and Hanchen Li and Urmish Thakker and James Zou and Kunle Olukotun},
      year={2026},
      eprint={2510.04618},
      archivePrefix={arXiv},
      primaryClass={cs.LG},
      url={https://arxiv.org/abs/2510.04618}, 
}

@misc{nanjundappa2025branch,
      title={Context Branching for LLM Conversations: A Version Control Approach to Exploratory Programming}, 
      author={Bhargav Chickmagalur Nanjundappa and Spandan Maaheshwari},
      year={2025},
      eprint={2512.13914},
      archivePrefix={arXiv},
      primaryClass={cs.SE},
      url={https://arxiv.org/abs/2512.13914}, 
}

@article{vasilopoulos2026codified,
  author  = {Vasilopoulos, Alexandros},
  title   = {Codified Context: Infrastructure for {AI} Agents
             in a Complex Codebase},
  journal = {arXiv preprint arXiv:2602.20478},
  year    = {2026},
  url     = {https://arxiv.org/abs/2602.20478}
}

@misc{wu2026gitcontextcontrollermanage,
      title={Git Context Controller: Manage the Context of LLM-based Agents like Git}, 
      author={Junde Wu and Minhao Hu and Jiayuan Zhu and Jiazhen Pan and Yuyuan Liu and Min Xu and Yueming Jin},
      year={2026},
      eprint={2508.00031},
      archivePrefix={arXiv},
      primaryClass={cs.SE},
      url={https://arxiv.org/abs/2508.00031}, 
}

@misc{navid2025compaction,
  author       = {P. Navid},
  title        = {Automatic Context Compaction for Agentic Workflows},
  howpublished = {Anthropic Cookbook,
                  \url{https://platform.claude.com/cookbook/tool-use-automatic-context-compaction}},
  month        = nov,
  year         = {2025},
  note         = {[Online; accessed March 2026]}
}

@misc{gemmateam2024gemma2,
      title={Gemma 2: Improving Open Language Models at a Practical Size}, 
      author={Gemma Team},
      year={2024},
      eprint={2408.00118},
      archivePrefix={arXiv},
      primaryClass={cs.CL},
      url={https://arxiv.org/abs/2408.00118}, 
}

@misc{robbes2026agentic,
      title={Agentic Much? Adoption of Coding Agents on GitHub}, 
      author={Romain Robbes and Théo Matricon and Thomas Degueule and Andre Hora and Stefano Zacchiroli},
      year={2026},
      eprint={2601.18341},
      archivePrefix={arXiv},
      primaryClass={cs.SE},
      url={https://arxiv.org/abs/2601.18341}, 
}

@inproceedings{wilhelm2025energypertoken,
  author    = {Patrick Wilhelm and Thorsten Wittkopp and Odej Kao},
  title     = {Beyond Test-Time Compute Strategies: Advocating Energy-per-Token in LLM Inference},
  booktitle = {Proceedings of the 5th Workshop on Machine Learning and Systems (EuroMLSys '25)},
  year      = {2025},
  address   = {Rotterdam, Netherlands},
  publisher = {ACM},
  doi       = {10.1145/3721146.3721953},
  url       = {https://euromlsys.eu/pdf/euromlsys25-27.pdf}
}

@misc{husom2024profiling,
      title={The Price of Prompting: Profiling Energy Use in Large Language Models Inference}, 
      author={Erik Johannes Husom and Arda Goknil and Lwin Khin Shar and Sagar Sen},
      year={2026},
      eprint={2407.16893},
      archivePrefix={arXiv},
      primaryClass={cs.CY},
      url={https://arxiv.org/abs/2407.16893}, 
}

@inproceedings{li2024sprout,
    title = "Sprout: Green Generative {AI} with Carbon-Efficient {LLM} Inference",
    author = "Li, Baolin  and
      Jiang, Yankai  and
      Gadepally, Vijay  and
      Tiwari, Devesh",
    editor = "Al-Onaizan, Yaser  and
      Bansal, Mohit  and
      Chen, Yun-Nung",
    booktitle = "Proceedings of the 2024 Conference on Empirical Methods in Natural Language Processing",
    month = nov,
    year = "2024",
    address = "Miami, Florida, USA",
    publisher = "Association for Computational Linguistics",
    url = "https://aclanthology.org/2024.emnlp-main.1215/",
    doi = "10.18653/v1/2024.emnlp-main.1215",
    pages = "21799--21813"
}

@misc{vercel2024nextconfig,
  author       = {{Vercel}},
  title        = {next.config.js {Configuration} -- {Next.js} Documentation},
  howpublished = {\url{https://nextjs.org/docs/app/api-reference/config/next-config-js}},
  year         = {2024},
  note         = {[Online; accessed March 2026]}
}

@misc{anthropic2025mcpjson,
  author       = {{Anthropic}},
  title        = {Model {Context} {Protocol} ({MCP}) in {Claude} {Code}},
  howpublished = {\url{https://docs.anthropic.com/en/docs/claude-code/mcp}},
  year         = {2025},
  note         = {[Online; accessed March 2026]}
}

@article{perera2025acm,
  author    = {Manoj Madushanka Perera and Adnan Mahmood and Kasun Eranda Wijethilake and Quan Z. Sheng},
  title     = {Towards Adaptive Context Management for Intelligent Conversational Question Answering},
  journal   = {arXiv preprint arXiv:2509.17829},
  year      = {2025},
  url       = {https://arxiv.org/abs/2509.17829},
  note      = {%% VERIFIED: arXiv:2509.17829 confirmed from tokalator.wiki/wiki catalog.
               %% Note: existing zhang2025adaptive key may refer to a different paper;
               %% this entry is the wiki-confirmed ACM paper.},
}

@misc{quantization,
      title={Understanding Efficiency: Quantization, Batching, and Serving Strategies in LLM Energy Use}, 
      author={Julien Delavande and Regis Pierrard and Sasha Luccioni},
      year={2026},
      eprint={2601.22362},
      archivePrefix={arXiv},
      primaryClass={cs.LG},
      url={https://arxiv.org/abs/2601.22362}, 
}

@misc{google2026genaisdk,
  author = {{Google DeepMind}},
  title = {Google Gen AI SDK for Python},
  year = {2026},
  publisher = {GitHub},
  journal = {GitHub repository},
  howpublished = {\url{https://github.com/googleapis/python-genai}},
  note = {Official Python client for the Gemini API and Vertex AI}
}

\end{document}